\documentclass[useAMS,usenatbib]{mn2e}
\usepackage{graphicx}
\usepackage{amssymb, amsmath}
\usepackage{array}

\def\Sun{{\sun}}
\def\aj{{\it AJ}}
\def\aap{{\it Ast.~\&Astrophys.}}
\def\apj{{\it Ap.~J.}}
\def\apjl{{\it Ap.~J.~Lett.}}
\def\nat{{\it Nature}}

\def\mnras{{\it Mon}. {\it Not}. {\it R}. {\it Astr}. {\it Soc}.,~}
\def\Earth{\earth}

\def\MJup{$\mathrm{M_{Jup}}$}

\def\fenh{\ensuremath{f_{\mathrm{enh}}}}
\def\taudecay{\ensuremath{\tau_{\mathrm{decay}}}}
\newcommand{\ME}{\ensuremath{\mathrm{M_{\Earth}}}}
\newcommand{\MSun}{\ensuremath{\mathrm{M_{\odot}}}}
\newcommand{\Mtot}{\ensuremath{\mathrm{M_{\mathrm{tot}}}}}
\newcommand{\mtot}{\ensuremath{\mathrm{m_{\mathrm{tot}}}}}

\newcolumntype{x}[1]{%
>{\raggedleft\hspace{0pt}}p{#1}}%
\newcommand{\tn}{\tabularnewline}

\newcommand{\simgeq} {\mbox{\raisebox{-0.5ex}{$\textstyle \sim$} \raisebox{ 0.8ex}{$\textstyle  \!\!\!\!\!\! >$  }}}

\title[On the formation of hot Neptunes and super-Earths]{On the
  formation of hot Neptunes and super-Earths} \author[D.~S.~McNeil and
  R.~P.~Nelson]{D.~S.~McNeil$^{1}$\thanks{E-mail: d.mcneil@qmul.ac.uk}
  and R.~P.~Nelson$^{1}$\\ $^{1}$Astronomy Unit, School of
  Mathematical Sciences, Queen Mary University of London, Mile End
  Road, London, UK E1 4NS}

\begin{document}

\date{Accepted 2009 September 30.  Received 2009 September 7; in original form 2009 June 22}

\pagerange{\pageref{firstpage}--\pageref{lastpage}} \pubyear{2009}

\maketitle

\label{firstpage}

\begin{abstract}
  
  { The discovery of short-period Neptune-mass objects, now including
    the remarkable system HD69830 \cite{lovis06} with three Neptune
    analogues, raises difficult questions about current formation
    models which may require a global treatment of the protoplanetary
    disc.
    Several formation scenarios have been
    proposed, where most combine the canonical oligarchic picture of
    core accretion with type I migration (e.g.~\citealt*{terq}) and
    planetary atmosphere physics (e.g.~\citealt{alib}).  To date, due
    in part to the computational challenges involved, published
    studies have considered only a very small number of progenitors at
    late times.  This leaves unaddressed important questions about the
    global viability of the models.  We seek to determine whether the
    most natural model -- namely, taking the canonical oligarchic
    picture of core accretion and introducing type I migration -- can
    succeed in forming objects of 10 Earth masses and more in the
    innermost parts of the disc. }
  
  
  { This problem is investigated using both traditional semianalytic
    methods for modelling oligarchic growth as well as a new parallel
    multi-zone N-body code designed specifically for treating
    planetary formation problems with large dynamic range
    \citep{mcn09}.}
  { We find that it is extremely difficult for oligarchic tidal
    migration models to reproduce the observed distribution.  Even
    under many variations of the typical parameters, including cases
    in which after the amount of mass in our disc is greatly increased
    above the standard Hayashi minimum-mass model, we form no objects
    of mass greater than 8 Earth masses.  By comparison, it is
    relatively straightforward to form icy super-Earths.}
    
    
    {We conclude that either the initial conditions of the
      protoplanetary discs in short-period Neptune systems were substantially
      different from the standard disc models we used, or there is
      important physics yet to be understood and included in models of
      the type we have presented here.}
    
\end{abstract}

\begin{keywords}
  planetary systems: formation
\end{keywords}

\section{Introduction}
\label{section:intro}

At present, there are $\sim\!$19 known extrasolar
planets\footnote{Data taken from the Extrasolar Planets Encyclopedia,
  Schneider, J., http://exoplanet.eu} with estimated masses between
0.03 and 0.12 \MJup, or between $\sim\!10$ and $\sim\!40$ \ME.  With
one exception -- OGLE-05-169L b, at 2.8 AU -- all of the planets have
semimajor axes smaller than 1 AU, and so are reasonably called hot
Neptunes.  In fact, with only one more exception, HD69830 d, all have
semimajor axes $< 0.23$ AU.  To determine whether the objects are
genuine Neptunes, i.e.~ice giants, and not merely very large rocky
bodies better thought of as super-Earths, knowledge of the mean
densities is required.  Three of these objects have known radii, and
two have mean densities compatible with being a Neptune-like body.
GJ436 b \citep{butler2004} has a density of
$\sim\!1.69\,\mathrm{g/cm^3}$ \citep{torres}, and HAT-P-11 b
\citep{bakos} has a density of $\sim\!1.33\,\mathrm{g/cm^3}$; compare
Neptune with $1.64\,\mathrm{g/cm^3}$.  The HD69830 system
\citep{lovis06} is particularly interesting.  It contains three
Neptune-like objects: HD69830 b at 0.0785 AU, of 10.5 \ME$\,$ and
eccentricity 0.1; HD69380 c at 0.186 AU, of 12.1 \ME$\,$ and
eccentricity 0.13; and HD69380 d at 0.63 AU, of 18.4 \ME, and
eccentricity 0.07 ($\pm0.07$, so a near-circular orbit is possible).
The system was also reported to have a disc of warm infrared-emitting
dust located between planets c and d \citep{beichman2005}, presumably
due to a collisionally active remnant asteroid belt.  More recent work
\citep{lisse07} suggests instead that the infrared excess is due to a
debris disc outside the known planets at $\simeq 1$ AU, probably
resulting from the breakup of an asteroid.  This system clearly
provides a rigorous test of planet formation and migration theories.

Indeed, short-period Neptune-mass bodies have several advantages as
probes of planetary origins.  They are large enough to be observable,
but small enough that we need not consider gravitational disc
instability as a formation process.  They are also large enough to
undergo significant type I migration, but not so large that they can
open a gap in the disc and undergo type II migration \citep{pap84,
  bryden99, crida06}.  Accordingly, hot Neptunes can yield less
ambiguous tests of type I migration than more massive planets which
could have significantly perturbed the gas disc they were embedded in,
and their existence provides strong evidence that some kind of type I
migration is operating in protoplanetary discs.

Hot Neptunes are therefore useful for exploring the intersection of
the classical oligarchic core accretion picture (\citealt*{ko98};
e.g.~\citealt*{cham01, thommes03}) with type I migration
\citep{ward97}.  Two main classes of scenario exist in the literature:
one which concentrates on the atmospheric gas physics, and one which
concentrates on the dynamics of the protoplanetary interactions.

\cite{alib} incorporate sophisticated atmospheric physics and follow
the evolution of effectively isolated cores through the disc as they
grow via planetesimal accretion, migrate inwards due to type I
effects, accrete gas after the accretion rate of solids drops, and
finally have their atmospheric mass reduced after arriving in the
short-period region by evaporation.  However, the history for the
HD69830 system proposed in \cite{alib} is difficult to reconcile with
the oligarchic paradigm \citep{ko98}.  Their best-fitting model
involves exactly three seed objects of masses $M = 0.6\,\ME$ at
semimajor axes 3 AU, 6.5 AU, and 8 AU, which migrate to small
semimajor axis through a mostly pristine planetesimal disc and
therefore can accrete a fair amount of material \citep{growthofmig}.
Where did these three progenitor objects come from?  In an oligarchic
framework, accretion in a narrow region -- even in the presence of
type I drag (see \citealt{komI, mcn05, komII}) -- results in many
roughly equal-mass bodies separated by a distance of $10\sim20$ Hill
radii.  These bodies then interact and merge in the giant-impact phase
of planet formation, and consume the accessible remnant planetesimal
disc.  That is, if a half-Earth-mass seed can successfully form at 3
AU, there should be a large number of similar seeds of varying mass
inside, each of which will accrete and migrate in similar ways.
Moreover, the interior seeds are likely to have formed first.  The net
result is that any such inward-migrating seed should be migrating
through a region highly depleted by the previous seeds, and by the
time an 0.6 \ME$\,$ core is formed at 3 AU much of the interior disc
should have gone to completion, and possibly formed its own planets
\citep{cham08}. There is no obvious mechanism to suppress embryo
growth everywhere in the disc except at three specific locations.  An
additional problem is that simulations have demonstrated that cores
migrating through a planetesimal swarm are unable to grow at the rate
prescribed by the \cite{alib} model.  In the presence of gas drag,
mean motion resonances cause the majority of the interior
planetesimals to be shepherded rather than being accreted, resulting
in planets whose masses are too low, and in the wrong ratio, compared
to those observed in the HD69830 system \citep{payne2009}.

Though not aiming at HD69830 in particular, \cite{terq} study the
formation of hot super-Earths and Neptunes by following the evolution
of $10-25$ planets of $0.1$ or $1\,\ME$ placed interior to 2 AU under
type I drag, include tidal interactions with the star, and use an
inner cavity in the gas disc at $\sim\!0.05$ AU.  For various disc
parameters, they succeed in making several objects of mass $\geq 8
\,\ME$, with a maximum of 12 \ME.  They find that the typical result
has between 2--5 planets, usually on near-commensurable orbits (with
strict commensurability often being broken by tidal circularization).
It is not clear whether this will scale up to larger masses, as they
performed only one run with total mass larger than 12 \ME (namely 25
\ME), and so an HD69830-like system would not appear in their results
even if the model would have succeeded in producing them.  Very little
material was lost from their runs, suggesting it might be possible.
However, their initial configurations are difficult to reconcile with
an oligarchic migration process.

Scenarios involving oligarchic formation and type I migration do not
exhaust the possible formation histories of the low-mass hot exoplanet
population, although they are arguably the most natural.
\cite{raymond08} surveys several other proposed possibilities (for the
terrestrial-mass regime): in situ formation; shepherding by migrating
giant planets or secular resonance sweeping; tidal circularization;
and photoevaporation of giant planets.  We will concentrate instead on
the simplest oligarchic type I migration picture, which should have
more success self-consistently generating an icy planet population,
and attempt to determine whether the fiducial models can reproduce the
observed distribution of hot Neptunes.  If they can, all the better;
if they cannot, then the specifics of their failure may point in the
direction of a solution and help us choose between the various
possibilities on offer.

To understand how short-period Neptunes are formed, we need to move
toward self-consistent global N-body models such as those which have
proved useful in understanding the formation of the terrestrial
planets and the outer solar system.  One problem presents itself
immediately: the short-period exoplanet problem has a formation time
to dynamical time ratio $\sim\!30$ times larger than the equivalent
terrestrial formation problem, and $\sim\!1000$ times that of the
equivalent Jovian core formation problem.  The lifetime of the gas
disc (with a probable upper limit of $\sim6$ Myr) is a common
reference time-scale for each problem, as gas giants must form while
there is still gas for them to accrete, and even ice giants which
attain short-period orbits must have migrated while there is gas
present.  However, the characteristic orbital periods for each
formation problem vary from $0.01$ yr at 0.05 AU to $0.35$ yr at 0.5
AU to $11$ yr at 5.0 AU.  (Indeed, $0.01$ yr is optimistic, as many
exoplanets are closer to their parent stars than 0.05 AU.)  Since most
state-of-the-art planetary N-body codes require the integration
timestep to be some small fraction of the orbital period of the
innermost object, preserving the same wall-clock run times that
researchers have become accustomed to would usually require reducing
the number of planetesimals in a simulation by orders of magnitude,
producing an unacceptable decrease in resolution.

It is extremely unlikely that this difficulty can be eliminated
entirely, as it is a consequence of the strong dependence of orbital
period on semimajor axis in the Kepler problem.  That said, the
challenge can be managed to some extent by taking advantage of the
scale separation caused by the troublesome dynamic range.  The
standard approaches use a common drift timestep for all particles (and
therefore ``over-integrate'' the more distant objects) and compute the
(non-encountering) forces between the particles at the same frequency,
both of which involve far more computation on the distant, slow-moving
objects than is necessary to preserve qualitatively accurate dynamics.
In \cite{mcn09} the authors combine various techniques in the
literature (\citealt*{dll98, cham99, saha92}) to construct a new
algorithm which allows for radial zones with different timesteps and
different inter-zone force evaluation frequencies, but reduces to the
proven techniques of \cite{dll98} and \cite{cham99} for objects within
the same zone.  This allows new trade-offs between force accuracy and
run time.

In our first paper, \cite{mcn09}, we addressed the numerical
challenges of studying oligarchic models of short-period exoplanet
formation.  Here we apply a new code with a parallel implementation of
those methods, to determine the ``reference population'' of planets
predicted by the fiducial models.  We integrate global planetesimal
discs extending from 0.05 AU to 10 AU under various models of the
protoplanetary disc.  Our models, both semianalytic and N-body, are
described in \S\ref{modeldescr}, and the simulation conditions in
\S\ref{simcond}.  Simulation results are presented in \S\ref{results}
and discussed in \S\ref{discussion}, and we conclude in
\S\ref{section:conc}.

\section{Description of models}
\label{modeldescr}
  To model the formation of planets in a global disc with a large
  dynamic range, we will divide the system into three stages: (1) the
  first stage, corresponding to the first 0.4 Myr after our nominal
  starting time t=0, which we will model using a semianalytic
  treatment based upon that of \cite{thommes03}; (2) the middle stage,
  from t=0.4 Myr to 6 Myr, which we will model using the new
  multiscale N-body code; and (3) the late post-gas stage, from 6 Myr
  to 100 Myr, during which we will evolve the inner regions of the
  disc using a traditional SyMBA implementation.

  As usual, the semianalytic model serves two purposes.  It allows for
  rough exploration of parameter space and provides self-consistent
  initial conditions for the second stage.  In practice, we find that
  for the purposes of serving as initial conditions, the model is
  often more than is necessary: as long as the initial mass of the
  growing protoplanets is much less than their final mass, and they
  have many encounters before they undergo significant migration, the
  systems do not appear to show strong dependence on the details of
  the initial conditions.  As a quasi-equilibrium, oligarchy is quite
  robust.  Nevertheless, although using the model does not eliminate
  the problems caused by initial conditions with inconsistent
  histories, it does help mitigate them.
  
  To explain our approach, we will first introduce the general
  features of the gaseous and solid material of the protoplanetary
  disc in sections \ref{gasdisc} and \ref{solid}, and then describe
  the prescriptions for the aerodynamic and type I drags applied to
  objects by the gas disc in sections \ref{aero} and \ref{typeI}.
  These models are shared by both our semianalytic approach for the
  early stage, summarized in \ref{semi}, and our N-body approach to
  the later stages, summarized in \ref{nbody}.

  \subsection{Gas disc}
  \label{gasdisc}
  For simplicity we will consider only models resembling those of the
  minimum-mass solar nebula (MMSN) of \cite{hay1981}, in which all
  physical disc parameters may be expressed as simple functions of the
  cylindrical radius from the star $r$ and the height above the disc
  midplane $z$.

  We take the volume density of the gas to be
  \begin{equation}
    \rho_{\rm{gas}}(r,z) =
    \Sigma_{\rm{gas}} / \sqrt{2 \pi} z_0 \, \exp(-z^2/2 z_0^2)
    \end{equation} 
  where $(r,z)$ are cylindrical coordinates and $z_0$ is the disc
  thickness.  We set
  \begin{equation}
    \Sigma_{\rm{gas}} = \Sigma^g_{\rm{1 AU}} \, (r/{\mathrm{AU}})^{-\alpha}
    \end{equation} 
  where $\Sigma^g_{\rm{1 AU}}$ is the gas surface density at 1 AU
  ($1704 \, {\mathrm{g/cm^2}}$ in the MMSN) and $\alpha$ gives the
  radial dependence of the density ($\alpha=1.5$ in the MMSN).  In
  practice, the quantity of interest is usually the ratio of mass in
  the disc to the mass in the MMSN.  We label this ratio $\fenh$, and
  determine the appropriate $\Sigma^g_{\rm{1 AU}}$ by normalizing so
  that the amount of gas mass from 0.05 AU to 15 AU in each simulation
  is $\fenh$ times the MMSN value.  In general terms the disc masses
  that we adopt are consistent with disc masses inferred from sub-mm
  observations \citep{andrews2005, andrews2007}. More specifically,
  \cite{andrews2005} indicate that more than one third of discs in the
  Taurus-Auriga region have masses which exceed that of the MMSN
  model, with similar statistics applying to discs observed in the
  $\rho$ Ophiuchus complex \citep{andrews2007}.

  For the disc height we take
  \begin{equation}
    \label{eq:z0}
    z_0 = 0.0472 \, \mathrm{AU} \, ({r/{\mathrm{AU}}})^{5/4}
  \end{equation}
  Note that this power-law in $r$ is traditional in the N-body
  planetary formation community but differs from the canonical choice
  in the planetary hydrodynamics community of taking a constant
  $z_0/r$ ratio of 0.05 or 0.07. (Although the difference may appear
  minor, it changes the surface density power-law at which neighbouring
  equal-mass objects undergoing type I migration switch from
  convergent to divergent migration, and is known to be important when
  reconciling results from different simulations involving capture
  into resonance: see \citealt*{cresswell}.)  We assume
  the gas is in perfect cylindrical rotation.  We parametrize the
  pressure support by
  \begin{equation}
    \eta = 0.6 (z_0/a)^2
  \end{equation}
  with typical $\eta \simeq 0.001$, and take
  \begin{equation}
    v_{\mathrm{gas}} = v_{\mathrm{kep}} \sqrt{1 - 2 \eta}
    \end{equation}
  where $v_\mathrm{kep}$ is the orbital velocity for a circular orbit
  at $r$.

  The dissipation of the gas disc with time $t$ is treated by
  introducing a single uniform exponential damping time-scale,
  \begin{equation}
    \label{eq:decay}
    \rho_{\rm{gas}} \propto \exp(-t/\taudecay)
  \end{equation}
  following \cite{kom02}. We neglect the gravitational potential due
  to the disc.
  
  \subsection{Solid material}
  \label{solid}

  At time $t=0$, we set a fixed gas-to-rock ratio of
  $\Sigma_{\rm{gas}} = 240 \, \Sigma_{\rm{rock}}$ (from the classical
  values of $\sim\!7\, {\mathrm{g/cm^2}}$ for solids at 1 AU and
  $\sim\!1700\, {\mathrm{g/cm^2}}$ for the gas, \citealt*{hay1981}), and
  therefore take the initial surface density in rocky material to be
  of the form
  \begin{equation}
    \Sigma_{\rm{rock}} = \Sigma^r_{1 \rm{AU}} ({r}/{\rm{AU}})^{-\alpha}
  \end{equation}
  where $r$ is the cylindrical radius and $\Sigma^r_{1 \rm{AU}}$ is the
  surface density of solids at 1 AU.  After \cite{thommes03}, we
  introduce a smoothed snow line beyond which the amount of material
  in solids is enhanced due to the temperature decreasing sufficiently
  to allow the condensation of ices.  Placing the snow line at
  $S_{\mathrm{loc}}$ and using a smoothing scale for the transition of
  $S_{\mathrm{sm}}$ = 0.25 AU, and using an enhancement factor
  $S_{\mathrm{enh}}$, we write
  \begin{equation}
    S = (0.5 \tanh( (r - S_{\mathrm{loc}})/ S_{\mathrm{sm}}) + 0.5)
  \end{equation}
  so that the total initial surface density in solids
  \begin{equation}
    \Sigma_{\mathrm{solid}} = \Sigma_{\rm{rock}} (1-S) + S_{\mathrm{enh}} \Sigma_{\rm{rock}} (S) 
  \end{equation}
  The form of $S$ is unimportant, and is chosen simply to soften the
  discontinuity for numerical purposes.
  
  We assume that all solid material has a density of 2.0 ${\mathrm{g/cm^3}}$.
  
  \subsection{Aerodynamic drag}
  \label{aero}
  We will apply an aerodynamic drag to all bodies (although the effects
  will usually be negligible on objects larger than 1000 km).  We take
  the drag time-scale as
  \begin{equation}
    \label{eq:tau_aero_1}
    \tau_{\rm{aero}} = \frac{8}{3} \frac{\rho_m}{\rho_{\rm{gas}}}
    \frac{r_m}{C_D} \frac{1}{v_{\rm{rel}}}
  \end{equation}
  where $\rho_m$ is the mass density of an object, $r_m$ its radius,
  $v_{\rm{rel}}$ its velocity relative to the gas velocity at its
  position, and for simplicity we use a drag efficiency of $C_D = 1$.
  In our semianalytic model, we will use the approximately equivalent
  form
  \begin{equation}
    \label{eq:tau_aero_2}
    \tau_{\rm{aero}} = \frac{1}{e_m} \frac{m}{(C_D/2) \, \pi r_m^2 \,
      \rho_{\rm{gas}} \, a \, \Omega}
  \end{equation}
  where $m$ is the planetesimal's mass, $a$ its semimajor axis, $e_m$
  its eccentricity, and $\Omega$ its orbital frequency.
  
  Orbit-averaging the above expression to determine the planetesimal
  migration rate $v_m$ in the small $e$, $i$ (inclination), and $\eta$
  limit, \cite{adachi} (after correction by \citealt*{kary}) find
  \begin{eqnarray}
    \label{eq:v_aero}
    v_m = \frac{da}{dt} \! \Big\arrowvert_{\!\rm{aero}} \simeq
    - 2 \frac{a}{\tau_{\rm{aero}} \, e_m} \left( \frac{5}{8} e_m^2 +
    \frac{1}{2} i_m^2 + \eta^2 \right)^{1/2} \nonumber \\
    \bigg\{ \eta + \left( \frac{\alpha}{4} + \frac{5}{16} \right) e_m^2 + \frac{1}{8}
    i_m^2 \bigg\}
  \end{eqnarray}

  For the N-body code, we will use the acceleration
  \begin{equation}
    {\boldsymbol{a}}_{\mathrm{aero}} = \frac{d\boldsymbol{v}}{dt} = 
    \frac{\boldsymbol{v} - \boldsymbol{v}_{\mathrm{gas}}}{\tau_{\mathrm{aero}}}
  \end{equation}
  with $\tau_{\mathrm{aero}}$ from eq.~\ref{eq:tau_aero_1}.

  \subsection{Type I migration}
  \label{typeI}

  For semianalytic purposes, we use the type I migration equation of
  \cite{ttw}, which gives an orbit-averaged migration rate
  $dr/dt$ of
  \begin{equation}
    \label{eq:v_M}
    v_M = - c_a (2.7 + 1.1 \alpha) \left(\frac{M}{M_\Sun} \right) \left(
    \frac{\Sigma_{\mathrm{gas}} \, r^2}{M_\Sun} \right)
    \left(\frac{r}{z_0}\right)^2 r \, \Omega
  \end{equation}
  for an object of mass $M$ at distance $r$ with orbital frequency
  $\Omega$, where we introduce $c_a$ as a parameter to incorporate
  uncertainty in the migration efficiency.  We note that significant
  contributions to the corotation torque may arise due to non-linear
  effects, resulting in a significantly reduced migration rate
  \citep{masset2006, paardekooper2009}, such that $c_a < 1$.  The
  semianalytic model will assume that embryos are always on circular
  orbits and so we do not need an expression for the eccentricity
  damping.
  
  For N-body calculations, we will use the full instantaneous specific
  force expressions of \cite{ttw} and \cite{tw04}, as given in
  appendix A of \cite{komII}, which we do not repeat here.  Our only
  modification is to introduce a $c_a$ parameter into the relevant
  radial migration term of the form of eq.~\ref{eq:v_M}.

  There have been a number of developments over recent years which
  have led to modifications of the basic picture of how type I
  migration may operate under differing assumptions about the
  underlying protoplanetary disc structure. In addition to the
  corotation torque effects mentioned above, it has been noted that
  regions in the disc where the surface density profile has a positive
  gradient (such that the surface density increases outward) may act
  as planet traps, where planetary migration may not occur at all
  \citep{masset2006b}. A planet trap may exist near to the star where
  the stellar magnetosphere clears a low density cavity in the disc
  \citep{linbod}. As discussed in the introduction to this paper, and
  in section~\ref{comparison}, other researchers have considered the
  role of this cavity in the formation of short-period Neptune and
  super-Earth planets, and not surprisingly have found that the
  accumulation of planetary embryos there can lead to the formation of
  planets of the required mass.  However, the formation of multiple
  Neptune and/or super-Earth-mass planet systems with a broad range of
  semimajor axes, such as HD69830, do not arise naturally from these
  models.  Formation of a multiple planet system inside the cavity is
  possible, but the gravitational potential well of the star likely
  prevents scattering out to orbital radii that would be required to
  explain HD69830. \cite{morbidelli2008} examined the possibility of
  planetary growth at a planet trap located near the snowline.
  Probably the most plausible explanation for a planet trap being
  located there is the model of inside-out disc clearance by
  magnetohydrodynamic turbulence suggested by \cite{chiang2007}. The
  simulations by \cite{morbidelli2008} indicate that multiple planets
  at a planet trap located at the snowline can indeed undergo close
  encounters and collisions through mutual scattering, allowing large
  bodies to grow there. In addition, scattering out of the trap was
  observed, allowing planets to migrate inward after growth through
  tidal interaction with the interior disc.  One uncertainty which
  remains in this picture is how clear of material the inner disc
  needs to be for the inside-out clearing model of \cite{chiang2007}
  to operate.  X-rays generated in the stellar corona need to be able
  to penetrate deep into the disc and out to large radii for the
  planet trap to be located at the snowline, leading to uncertainty
  about how much type I migration can actually ensue once planets are
  scattered out of the trap. Given this uncertainty in the model, we
  have chosen to focus on the most generic type I migration scenario
  in the present study, which does not include planet traps located at
  specific radii.

  \subsection{Semianalytic approach}
  \label{semi}
  We follow \cite{mcn05}, which is derived from \cite{thommes03},
  replacing only the formula for type I drag used there (due to
  \citealt*{paplar}) with that of Tanaka and Ward (eq.~\ref{eq:v_M}).
  We will forego repeating the derivations and simply describe the
  resulting equations.

  We simulate the oligarchic migration formation scenario using a
  continuous two-component model consisting of protoplanetary embryos of
  mass $M(a)$ which accrete mass from a planetesimal field
  distribution with surface density $\Sigma_m(a)$ orbiting a star of
  mass $M_\Sun$, with gravitational constant $G$.  The embryos (of
  density $\rho_M$) are assumed to be kept at a constant fixed
  separation $b$ in single-planet Hill units ($r_{\mathrm{H}} = (M/3
  M_\Sun)^{1/3} a$) as a consequence of the usual oligarchic
  equilibration between the increased separation due to scattering and
  the decreased separation due to accretion (\citealt*{ko98}).  (As
  discussed in \citealt*{mcn05}, this approximation is questionable at
  late times when strong migration is operating or the discs are very
  massive, but is reasonable during the early stages.)  The embryo
  eccentricities are neglected as they are likely to be much smaller
  than those of the planetesimals due to both dynamical friction and
  type I damping (itself a kind of dynamical friction with the gas).
  The planetesimals (of density $\rho_m$) are assumed to have a
  uniform mass $m$ (and radius $r_m$), and at a given semimajor axis
  all have an equilibrium eccentricity found by equating the time-scale
  for stirring by the embryos by damping by aerodynamic drag.  See
  \cite{cham06} for a detailed description of the weaknesses of
  the equilibrium eccentricity assumption, and also note that by
  neglecting the contribution of embryo-embryo mergers we
  underestimate the accretion rate.

  The physics in the model is simple, although the algebra is somewhat
  tedious.  Embryos accrete mass from the planetesimals according to
  the expression\footnote{The factor of $b^{2/5}$ is missing from
    eq.~4 of \cite{mcn05}; the error was typographical.}
  \begin{equation}
    \label{eq:dMdt}
    \frac{dM}{dt}\Big\vert_{\rm{accr}} = \frac{3.93 \, M_\Sun^{1/6} \, G^{1/2} \,
      \Sigma_{\rm{m}} \, M^{2/3} \, b^{2/5} \, C_D^{2/5} \, \rho_{\rm{gas}}^{2/5}}
         {\rho_{M}^{1/3} \, a^{1/10} \, m^{2/15} \, \rho_{m}^{4/15}}
  \end{equation}
  which corresponds to a decrease in planetesimal surface density
  \begin{equation}
    \label{eq:dsdt}
    \frac{d\Sigma_{m}}{dt}\Big\vert_{\rm{accr}} = \frac{ -M_\Sun^{1/3} }{3^{2/3} \, b \, \pi
      a^2 \, M^{1/3}} \frac{dM}{dt}
  \end{equation}
  Planetesimals migrate due to aerodynamic drag with a radial rate
  $v_m$ given by eq.~\ref{eq:v_aero}, and embryos migrate due to type
  I effects with a radial rate $v_M$ given by eq.~\ref{eq:v_M}.  These
  four equations -- \ref{eq:dMdt}, \ref{eq:dsdt}, \ref{eq:v_aero}, and
  \ref{eq:v_M} -- are numerically integrated.

  We will assume that the separation $b=10$ and that all planetesimals
  have $r_m$ = 10 km.  We will take the initial seed mass for the
  embryos as $M(a,t=0) = 1.5 \times 10^{-3} M_{\Earth}$, which has the effect
  of artificially accelerating the growth of the more distant embryos
  during the earliest stages.
  
  \subsection{N-body approach}
  \label{nbody}
  The semianalytic Eulerian approach of the previous section can be
  useful as a rough estimate of the behaviour of the system, but
  contains very limited information about the dynamics involved.  For
  later stages in which the interactions between the embryos are
  significant a particle-based Lagrangian approach using an N-body
  code is necessary.

  Currently, the most robust N-body algorithm for oligarchic
  simulations of planet formation on long time-scales remains SyMBA
  \citep{dll98}, which derives from the original mixed-variable
  symplectic integrators of \cite{wishol} and \cite{kyn} but has been
  improved to treat close encounters between particles.
  Unfortunately, it requires a common base timestep for all particles,
  which is set to some small fraction (typically $\sim\!1/15$ or less)
  of the innermost period so that pericentre passage is always
  resolved.  (\citealt*{ld00} introduced a variant which could
  handle occasional objects crossing the usual innermost boundary by
  smoothly switching to a Bulirsch-Stoer integrator, following up on
  the innovations of Chambers 1999, but it becomes impractical when
  boundary crossings are common.)  In a companion paper \citep{mcn09}
  we introduce a new algorithm NAOKO which allows for multiple radial
  zones with distinct timesteps and can vary the number of force
  evaluations between different zones, making possible a new trade-off
  between the force accuracy between distant objects and speed.  We
  have implemented a parallel version of NAOKO in the planet formation
  code MIRANDA, which is basically a parallel SyMBA implementation.

  As in the semianalytics, we have two classes of objects, embryos and
  planetesimals, where the embryos can merge with each other and with
  the planetesimals, but the planetesimals do not self-interact either
  gravitationally or collisionally.  Two objects (assumed spherical)
  merge if their physical separation is less than the sum of their
  physical radii, and form one new object, conserving mass and linear
  momentum.  No fragmentation is considered.  Following standard
  practice (e.g.~\citealt*{thommes03}) we use ``super-planetesimals'',
  and replace the planetesimals of mass $m$ we would prefer to use
  with larger objects of mass $m_{\mathrm{sp}}$ which behave as
  objects of mass $m$ with respect to non-gravitational interactions
  like gas drag, and serve as representative elements of an underlying
  planetesimal population.  As long as the number of
  super-planetesimals is large and the embryo mass is greater than the
  super-planetesimal mass ($M \gg m_{\mathrm{sp}}$), the accretion
  rate is insensitive to this approximation.  (Taking a uniform $m$
  for the planetesimals and neglecting their self-interactions are
  considerably more damaging simplifications than using
  super-planetesimals in any event.)

  Interactions with the gas disc are limited to aerodynamic drag and
  type I drag, as described in sections \ref{aero} and \ref{typeI}.
  The super-planetesimals will experience an aerodynamic drag
  corresponding not to the physical radius of the integrated tracer,
  but to that of the underlying planetesimal (typically $\simeq 10$
  km).

  For the computationally challenging second stage during which the
  gas is present, we will integrate the system using MIRANDA in its
  new NAOKO mode \citep{mcn09}.  This allows the disc to be divided
  into distinct radial zones, and objects in each zone are integrated
  using different time steps.  In the simulations presented here we
  used four zones, with timesteps chosen so that all objects had at
  least $\simeq$ 15 steps per orbit: namely, $\Delta t$ = 0.53 yr for
  objects outside 4 AU; $\Delta t$ = 0.13 yr for 1.6-4 AU; $\Delta t$
  = 0.013 yr for 0.34-1.6 AU; and $\Delta t$ = 0.00083 yr for
  0.05-0.34 AU.  This corresponded to timestep ratios between outer
  and inner zones of 4, 10, and 16, respectively.  These ratios also
  describe the ratios of the frequencies on which the interzone forces
  were evaluated; intrazone forces were evaluated at each (zone) step.
  Boundaries between zones had associated transition zones centred
  there, of widths 0.5 AU, 0.1 AU, and 0.04 AU from outermost to
  innermost, in which the objects smoothly experienced both timesteps
  to avoid artificial kicks in velocity.  For the late stage after the
  gas disc has dissipated, the number of particles in the inner zone
  of interest will have decreased enough that we can return to the
  traditional SyMBA method to simulate the final giant-impact stage,
  with a fixed timestep of 0.0007 yr.

  \section{Simulation conditions}
  \label{simcond}

  Computational power being limited, it was not feasible even with the
  new code to perform multiple runs for each parameter set of
  interest.  Instead, we made compromises between coverage of
  parameter space and reproducibility of each run, and between
  concentrating on physically plausible scenarios and less plausible
  but informative limiting cases.  Table \ref{simparam} lists the
  resulting choices, and where two simulation labels are given we ran
  two instantiations which differed only in the random number seeds
  used to define the initial Keplerian angular variables of the
  particles.

  We used mass enhancement factors of $\fenh = 3, 5, 10$, to cover the
  range from the likely enhancements above the MMSN needed to form the
  solar system to something considerably larger.  (Again,
  $\Sigma^g_{\mathrm{1 AU}}$ is determined from normalizing the amount
  of mass between 0.05 AU and 15.0 AU to be $\fenh$ times the MMSN
  value.)  The surface density power-law was chosen from $\alpha =
  1.0, 0.5, 0.001$, so that the discs are all much flatter than the
  MMSN (with $\alpha=1.5$).  The MMSN value was skipped in the
  production runs presented here because preliminary low resolution
  simulations indicated that disc models with a high degree of central
  mass concentration do not successfully form surviving short-period
  planetary systems with significant amounts of mass. This may be due
  in part to the fact that such a disc model induces divergent type I
  migration for neighbouring bodies of equal mass, as discussed below,
  thereby reducing the planetary growth rate.  In addition, viscous
  disc models based on the \cite{shaksun} `alpha' prescription for
  viscosity tend to generate shallower surface density distributions
  \citep{papterq99, fogg07}.  Note that with $z_0/r \propto r^{1/4}$
  (eq.~\ref{eq:z0}), the type I $dr/dt$ rate is independent of $a$ for
  equal M at $\alpha=1.0$, and shows convergent migration for $\alpha
  < 1.0$.  We used disc dissipation time-scales (eq.~\ref{eq:decay})
  of 1 Myr and 2 Myr, which are roughly compatible with the observed
  disc decay times inferred from observations
  (e.g.~\citealt*{haisch01}).  The migration efficiency $c_a$ was set
  to either 0.3 or 1.0.  The snow line was placed at either
  $S_{\mathrm{loc}}$ = 2.7 AU or 4.0 AU, with an enhancement factor of
  $S_{\mathrm{enh}}=4$.  Some studies suggest that a snow line closer
  to the star (1.6--1.8 AU, \citealt*{lecar}) and a lower enhancement
  factor ($\sim\!2.2$, \citealt*{lodders}) may be more realistic, but
  others argue that a much higher enhancement can occur
  \citep{ciesla06}.  Since many of the Neptune-mass objects found to
  date orbit lower-luminosity stars, a closer-in snow line may more
  accurately model the environments of the known short-period Neptune
  planets.  However, to facilitate comparison with other work we will
  keep the mass of the central star at $M_\Sun$ and use the historical
  snow line.  (The $S_{\mathrm{loc}}$ = 4.0 AU runs were an experiment
  motivated by some preliminary simulations which suggested that
  delaying the onset of embryo growth past the snow line could help
  prevent promising objects from falling into the star;
  c.f. \citealt*{cham06}.)

  \begin{table}

      \caption{Simulation parameters.\label{simparam}}
      \begin{tabular}{@{}llllll@{}}
        \hline 
        Simulation & \fenh  & $S_{\mathrm{loc}}$ [AU] & $\alpha$ & $c_a$ &
        $\taudecay$ [Myr]\\
      
        \hline

        S01A, S01B & 3  & 2.7 & 0.001 & 0.3 & 1  \\
        S02A & 3  & 2.7 & 0.001 & 0.3 & 2  \\
        S03A, S03B & 3  & 2.7 & 0.001 & 1.0 & 1  \\
        S04A & 3  & 2.7 & 0.001 & 1.0 & 2  \\
        S05A, S05B& 3  & 2.7 & 0.5 & 0.3 & 1  \\
        S06A & 3  & 2.7 & 0.5 & 0.3 & 2  \\
        S07A, S07B & 3  & 2.7 & 0.5 & 1.0 & 1  \\
        S08A & 3  & 2.7 & 0.5 & 1.0 & 2  \\
        S09A, S09B & 3  & 2.7 & 1.0 & 0.3 & 1  \\
        S10A, S10B & 3  & 2.7 & 1.0 & 1.0 & 1  \\
        S11A & 3  & 2.7 & 1.0 & 1.0 & 2  \\
        S12A, S12B & 5  & 2.7 & 0.001 & 0.3 & 1  \\
        S13A & 5  & 2.7 & 0.001 & 0.3 & 2  \\
        S14A, S14B & 5  & 2.7 & 0.001 & 1.0 & 1  \\
        S15A & 5  & 2.7 & 0.001 & 1.0 & 2  \\
        S16A, S16B & 5  & 2.7 & 0.5 & 0.3 & 1  \\
        S17A, S17B & 5  & 2.7 & 0.5 & 1.0 & 1  \\
        S18A & 5  & 2.7 & 0.5 & 1.0 & 2  \\
        S19A, S19B & 5  & 2.7 & 1.0 & 1.0 & 1  \\
        S20A & 10 & 2.7 & 0.5 & 0.3 & 1  \\
        S21A & 10 & 2.7 & 0.5 & 1.0 & 1  \\
        S22A & 10 & 2.7 & 1.0 & 0.3 & 1  \\
        S23A & 3  & 4   & 0.001 & 0.3 & 1  \\
        S24A & 3  & 4   & 0.001 & 1.0 & 1  \\
        S25A & 3  & 4   & 0.5 & 0.3 & 1  \\
        S26A & 3  & 4   & 0.5 & 1.0 & 1  \\
        S27A & 3  & 4   & 1.0 & 0.3 & 1  \\
        S28A & 3  & 4   & 1.0 & 1.0 & 1  \\
        S29A & 5  & 4   & 0.001 & 0.3 & 1  \\
        S30A & 5  & 4   & 0.001 & 1.0 & 1  \\
        S31A & 5  & 4   & 0.5 & 0.3 & 1  \\
        S32A & 5  & 4   & 0.5 & 1.0 & 1  \\
        S33A & 5  & 4   & 1.0 & 0.3 & 1  \\
        S34A & 5  & 4   & 1.0 & 1.0 & 1  \\
        S35A & 10 & 4   & 0.5 & 1.0 & 1  \\
        S36A & 10 & 4   & 1.0 & 0.3 & 1  \\
        S37A & 10 & 4   & 1.0 & 1.0 & 1  \\
        
        \hline
      \end{tabular}

  \end{table}
  
  Each simulation was evolved for the first 0.4 Myr using the
  semianalytic model of \S\ref{semi}.  This yielded a distribution of
  embryo mass $M(a)$ and planetesimal surface density $\Sigma_m(a)$.
  These were then discretized into an N-body particle distribution
  extending from 1 AU to $\sim\!10$ AU, with spacing between the
  embryos fixed at $b\simeq10$ and the super-planetesimal mass
  $m_{\mathrm{sp}}$ chosen to ensure that locally $M/m_{\mathrm{sp}}
  \ge 5$ with $r_m$ = 10 km.  (This is a somewhat more relaxed
  condition than used in \cite{mcn05}, which required the relationship
  to hold globally and used one uniform $m_\mathrm{sp}$.)  The
  resulting particle sets had between 40 and 64 fully interacting
  embryos, with $>$ 10000 super-planetesimals for the main
  $S_{\mathrm{loc}}$ = 2.7 AU runs and $\sim\!4000$ super-planetesimals
  for the lower-resolution $S_{\mathrm{loc}}$ = 4.0 AU experiments.
  The initial embryo eccentricities and inclinations were arbitrarily
  set to 0.001 and 0.0005, respectively, and the planetesimal
  eccentricities to their (semianalytically-estimated) equilibrium
  values.  The simulations rapidly reach their random-velocity
  equilibrium.  The particle distributions were then integrated until
  t=6 Myr using the NAOKO algorithm of \S\ref{nbody}.  This 6 Myr
  time-scale was chosen so that even in the simulations with $\taudecay
  = 2$ Myr, at least 95\% of the original gas mass is gone.  Finally,
  the resulting protoplanets inside of 2 AU were then integrated until
  t=100 Myr using the traditional SyMBA algorithm.  (That is, the
  remnant planetesimal disc was removed entirely, as were all embryos
  with semimajor axis $>$ 2 AU.)

  Each integration took roughly $3\!\sim\!4$ weeks of runtime on an
  8-processor node for the 0.4-6 Myr phase, and then another week to
  two weeks in serial mode for the 6-100 Myr phase.  Without the use
  of both parallelism and the new NAOKO algorithm, the simulations
  would have taken an impractically long time.

  \section{Results}
  \label{results}

  In this section we present some sample runs in \S\ref{indruns},
  including both representative cases showing the radial evolution of
  the embryos in \S\ref{mighist} and the evolution of a few of our
  ``successful'' runs in \S\ref{S09a} and \S\ref{S16b}.  We discuss
  the global results in \S\ref{section:synthesis}, giving an overview
  in \S\ref{section:overview}.  The planetary mass distributions are
  covered in \S\ref{section:mass-dist}; the distribution of ices in
  \S\ref{section:ice-fraction}; various statistics on the resulting
  configurations in \S\ref{section:chambstat}; the final number of
  planets and the radial mixing in \S\ref{section:N}; the total
  surviving mass in the inner region in
  \S\ref{section:surviving_mass}; and, finally, the possibility of
  future mergers and the long-term stability of the systems in
  \S\ref{section:mergers}.

  \subsection{Description of individual runs}
  \label{indruns}
  
  \subsubsection{Sample migration histories}
  \label{mighist}

\begin{figure}

    \caption{Semimajor axis, perihelion, and aphelion for each embryo
      over time in simulation S14A.
      \label{S14A_qaQ}}
    \includegraphics[width=90mm, clip]{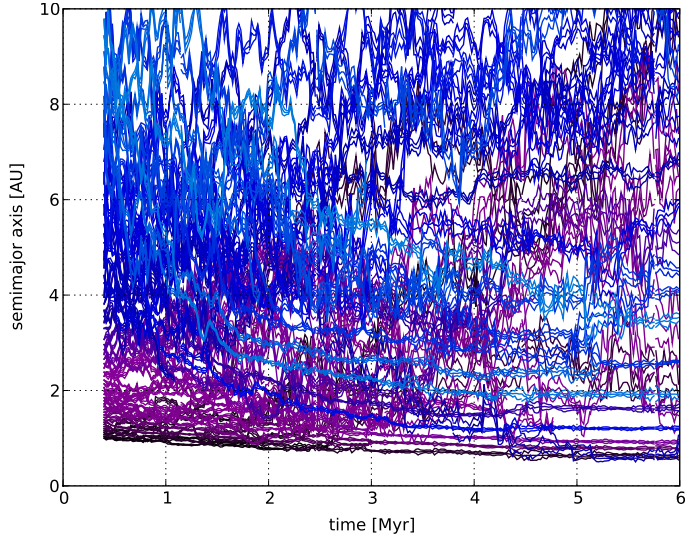}

  \end{figure}

\begin{figure}
     \caption{Semimajor axis, perihelion, and aphelion for each embryo
       over time in simulation S07A.
      \label{S07A_qaQ}}

   \includegraphics[width=90mm, clip]{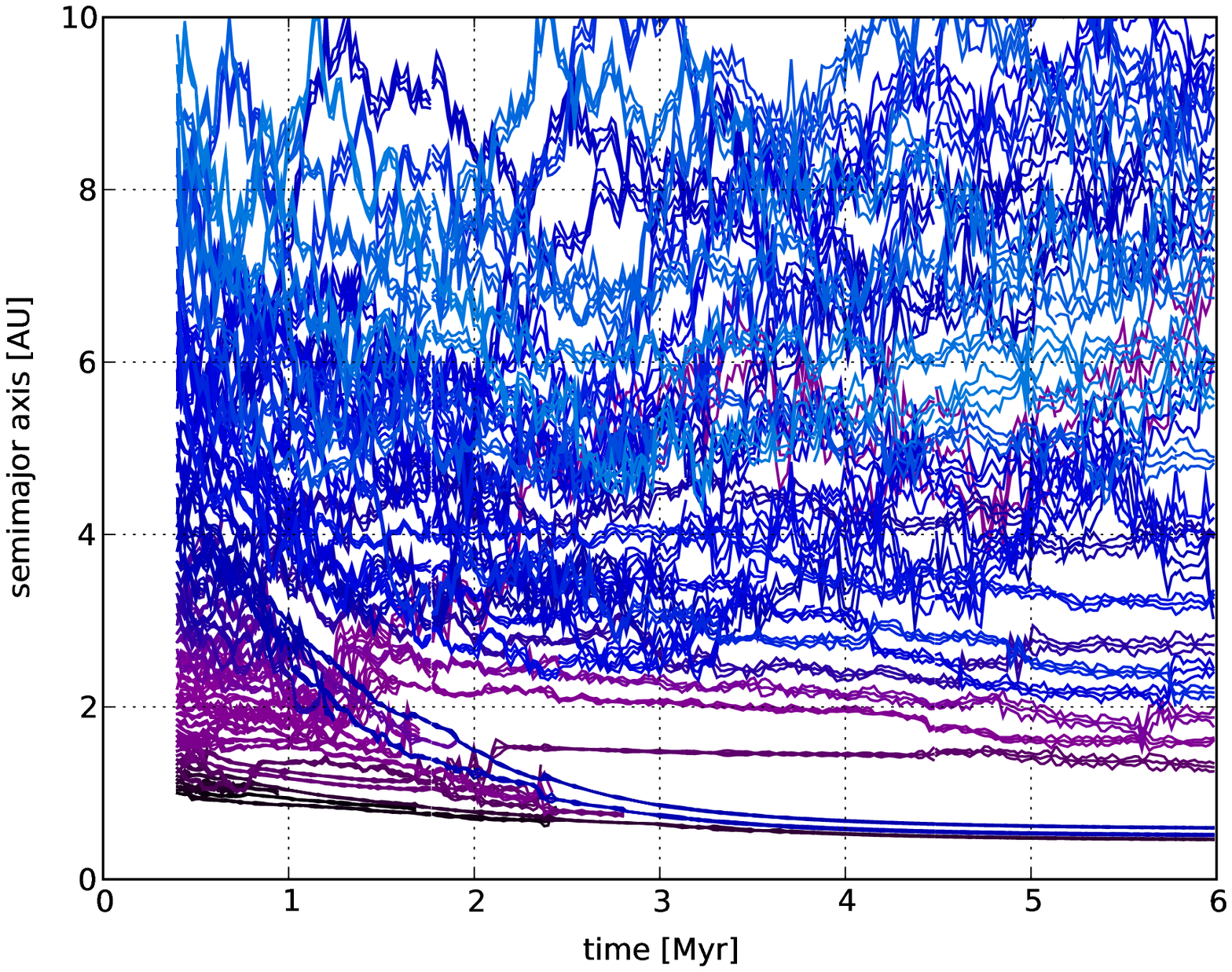}

  \end{figure}

\begin{figure}
      \caption{Semimajor axis, perihelion, and aphelion for each
        embryo over time in simulation S19A.
      \label{S19A_qaQ}}

    \includegraphics[width=90mm, clip]{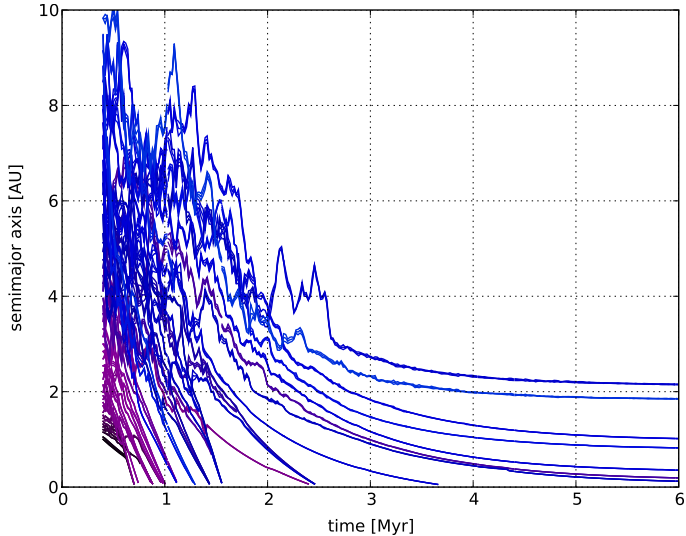}

  \end{figure}

  Figures \ref{S14A_qaQ}, \ref{S07A_qaQ}, and \ref{S19A_qaQ} offer
  example overviews of the radial evolution of the embryos.  In S14A,
  a flat disc ($\alpha=0.001$) with moderate enhancement $\fenh=5$,
  the outer regions are still chaotic at 6 Myr.  In S07A, a steeper
  disc with $\alpha=0.5$ and low enhancement $\fenh=3$, an object from
  beyond the snow line grows massive enough sufficiently early to
  migrate inwards much faster than its neighbours and compress the
  earlier objects into a much smaller radial region, causing both
  mergers and occasional outward ejections (e.g.~just after 2 Myr).
  There is little communication between the innermost and the
  outermost regions, creating a large empty region in $(a,t)$ space
  devoid of planetary embryos.  In S19A, $\alpha=1.0$ with moderate
  enhancement $\fenh=5$, we see that almost all of the inner material
  is rapidly lost to the inner edge of the simulation, and much of the
  material which eventually forms the resulting inner system comes
  from far beyond the snow line.  However, some of the resulting
  planets incorporate material which started inside of the snow line
  and was scattered out beyond it before migrating back in.  Each
  simulation had $c_a = 1.0$ and $\taudecay=$ 1 Myr, and therefore the
  majority of their tidal migration is complete by 3 Myr, although as
  S14A shows, a considerable amount of radial migration can continue
  due to embryo-embryo interactions long after the gas disc is gone.
  This can be compared with \cite{mcn05}'s study of terrestrial planet
  oligarchic migration scenarios, which predicted a tripartite
  division into an interior region of strong convoying behavior (where
  planets rapidly lock into mean-motion resonance and migrate in
  tandem), a transition region where objects `slide' towards their
  final destinations, and an outer region which remains chaotic.  In
  their study, they looked only at the $\alpha=1.5$ and $\alpha=1.0$
  cases, and did not consider any $\alpha$ generating convergent
  migration.

  \subsubsection{Run S09A}
  \label{S09a}
   The run S09A produced the single most massive object of any
   simulation, a planet of 7.2 \ME.  This particular run had the
   parameters $\fenh = 3, \alpha = 1.0, c_a = 0.3, \taudecay$ = 1 Myr,
   and $S_{\mathrm{loc}}$ = 2.7 AU.  At t=0.4 Myr, the simulation
   start, there were roughly comparable amounts of mass in embryos and
   planetesimals inside of 2 AU, \Mtot=1.1 \ME and \mtot=1.4 \ME,
   respectively.  By 1 Myr, both have increased, to \Mtot=2.3 \ME and
   \mtot=1.8 \ME.  The amount of embryo material in the region
   gradually increases to 11.1 \ME at 6 Myr.  This increase is
   non-monotonic, as embryos are occasionally lost in two ways.
   First, some are scattered beyond the inner simulation edge at 0.05
   AU where they are removed.  Second, some are scattered beyond the
   0.05-2 AU window, in which case they will remain in the system but
   no longer contribute to the mass (at least unless type I migration
   or embryo-embryo scattering brings them back in again).  By
   contrast, the planetesimal material reaches a maximum of $\simeq
   2.3 \ME$ at 2.2 Myr, as it migrates into the region via both Jacobi
   shepherding by embryos and aerodynamic drag, and then falls off to
   $\simeq 1.5 \ME$ for the remainder of the simulation, as material
   is consumed but new material is brought in to replace it.

  \begin{figure}
     \begin{center}
    \caption{Run S09A planets interior to 2 AU at t=6 Myr, planet mass
      versus semimajor axis, with bars indicating eccentric excursion.
      \label{plot:S09A_6Myr}
      }
    \includegraphics[width=90mm, clip]{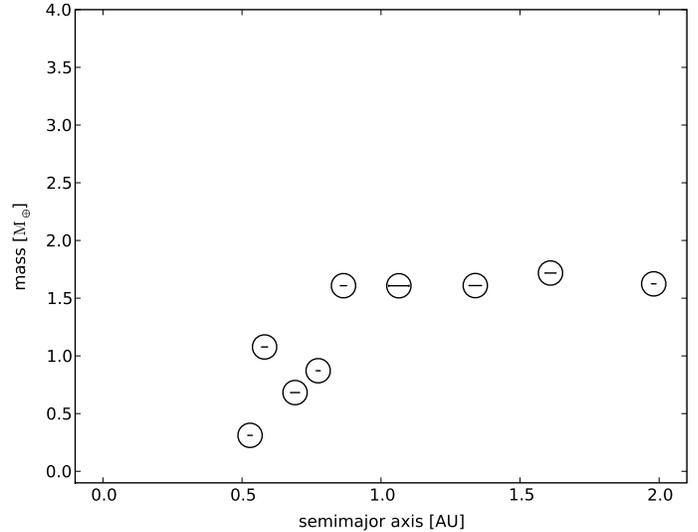}
    
     \end{center}
  \end{figure}

  \begin{figure}
     \begin{center}
       \caption{Run S09A planet mass versus time for all objects.
         From t=6 Myr, only the embryos interior to 2 AU are
         followed.
         \label{plot:S09A_Mvst}
         }
         \includegraphics[width=90mm, clip]{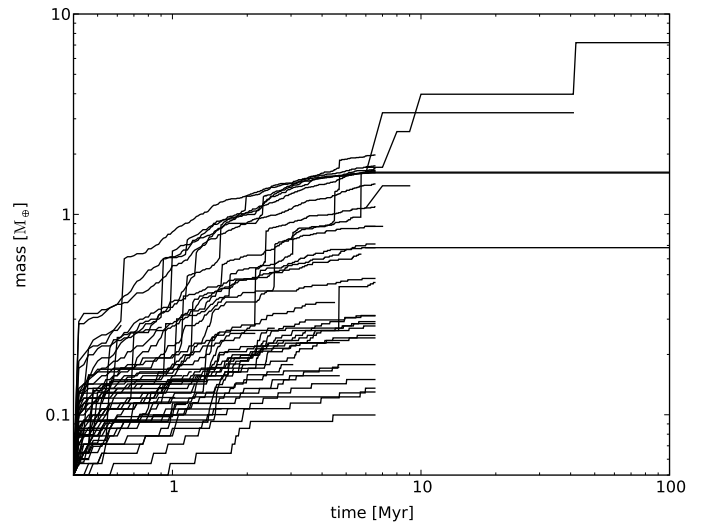}
     \end{center}
  \end{figure}

   As shown in figure \ref{plot:S09A_6Myr}, at 6 Myr there are nine
   embryos with $a <$ 2 AU, and they fall into two distinct
   categories: four objects inside of 0.8 AU, with masses less than
   1.2 \ME which vary by factors of three, and five well-spaced
   objects exterior to 0.8 AU which all have masses $\simeq1.6-1.7
   \ME$, where the inner three have startlingly similar masses, namely
   1.608 \ME, 1.609 \ME, and 1.607 \ME.  Figure \ref{plot:S09A_Mvst}
   shows that this is simply a coincidental result of equilibration
   processes, as the objects follow roughly similar growth curves --
   an early period of fast accretion transitioning to a much slower
   phase after 3 Myr when the gas is mostly gone -- but at times the
   objects differ in mass by a factor of $\simeq2$.  The three
   $\sim\!1.61 \ME\,$ objects also come from very different locations
   in the disc: the outermost embryo started its life at 2.4 AU, the
   middle at 3.3 AU, and the innermost at 1.6 AU.  During the giant
   impact phase, the objects at 0.87 and 1.06 AU merge quite quickly,
   at $\sim\!7$ Myr, as do the two innermost bodies.  By $\sim\!40$
   Myr, the original (t=6 Myr) 1.7 \ME body has consumed three more
   objects to reach its terminal mass, while the remaining objects
   were scattered outside 2 AU.

   \subsubsection{Run S16B}
   \label{S16b}

   \begin{figure}
     \begin{center}
      \caption{Run S16B planets at t=6 Myr.
         \label{plot:S16B_6Myr}
       }
       \includegraphics[width=90mm, clip]{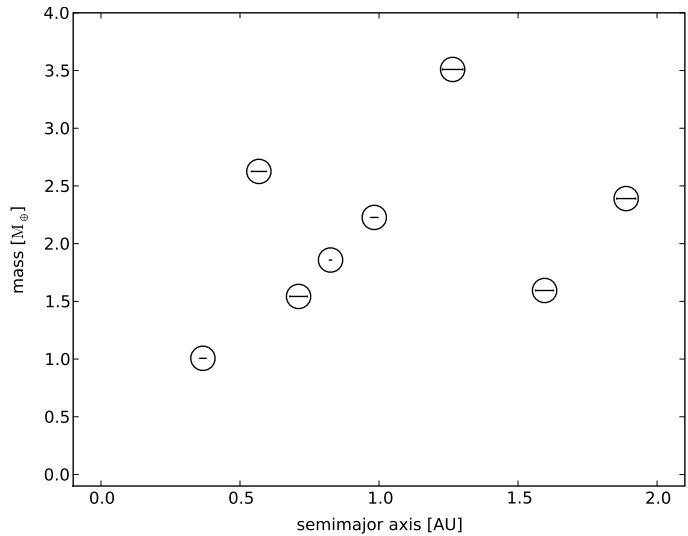}
     \end{center}
  \end{figure}

  The run S16B produced the largest amount of surviving embryo mass
  interior to 2 AU at 100 Myr of all the simulations, 16.75 \ME.  This
  particular run had the parameters $\fenh = 5, \alpha = 0.5, c_a =
  0.3, \taudecay$ = 1 Myr, and $S_{\mathrm{loc}}$ = 2.7 AU.  This
  region started with \Mtot=1.00 \ME and \mtot=1.3 \ME, reaching final
  values of \Mtot=16.8 \ME and \mtot=1.9 \ME at 6 Myr.

    As shown in figure \ref{plot:S16B_6Myr}, there are 8 embryos with
    at least an Earth mass of material (including a hot Earth of 1.01
    \ME at 0.37 AU).  The planet masses vary more than in S09A,
    although it is still true that the planets with the lowest masses
    are on the interior and the highest masses have semimajor axes
    between 1.2 and 1.6 AU.  The system is basically quiescent until a
    burst of activity between 14 and 15 Myr, during which the planet
    at 1 AU merges with the planets at 0.7 AU and 1.9 AU, and the
    planet at 0.6 AU merges with the planets at 0.8 AU and 1.6 AU.
    After this, the system does nothing but exchange angular momentum
    with little variation in semimajor axis.
    
    Comparing S09A with S16B suggests that maximizing the mass of the
    final planet does not require maximizing the mass of the planets
    when the giant impact phase begins; both simulations produced
    maximum objects of mass $6-7 \ME$, despite S16B having 1.5 times
    the mass of S09A at 6 Myr.

    \subsection{Synthesis of simulation outcomes}
    \label{section:synthesis}
    \subsubsection{Overview}
    \label{section:overview}

    \begin{figure*}
    \begin{center}
      \includegraphics[width=180mm, clip] {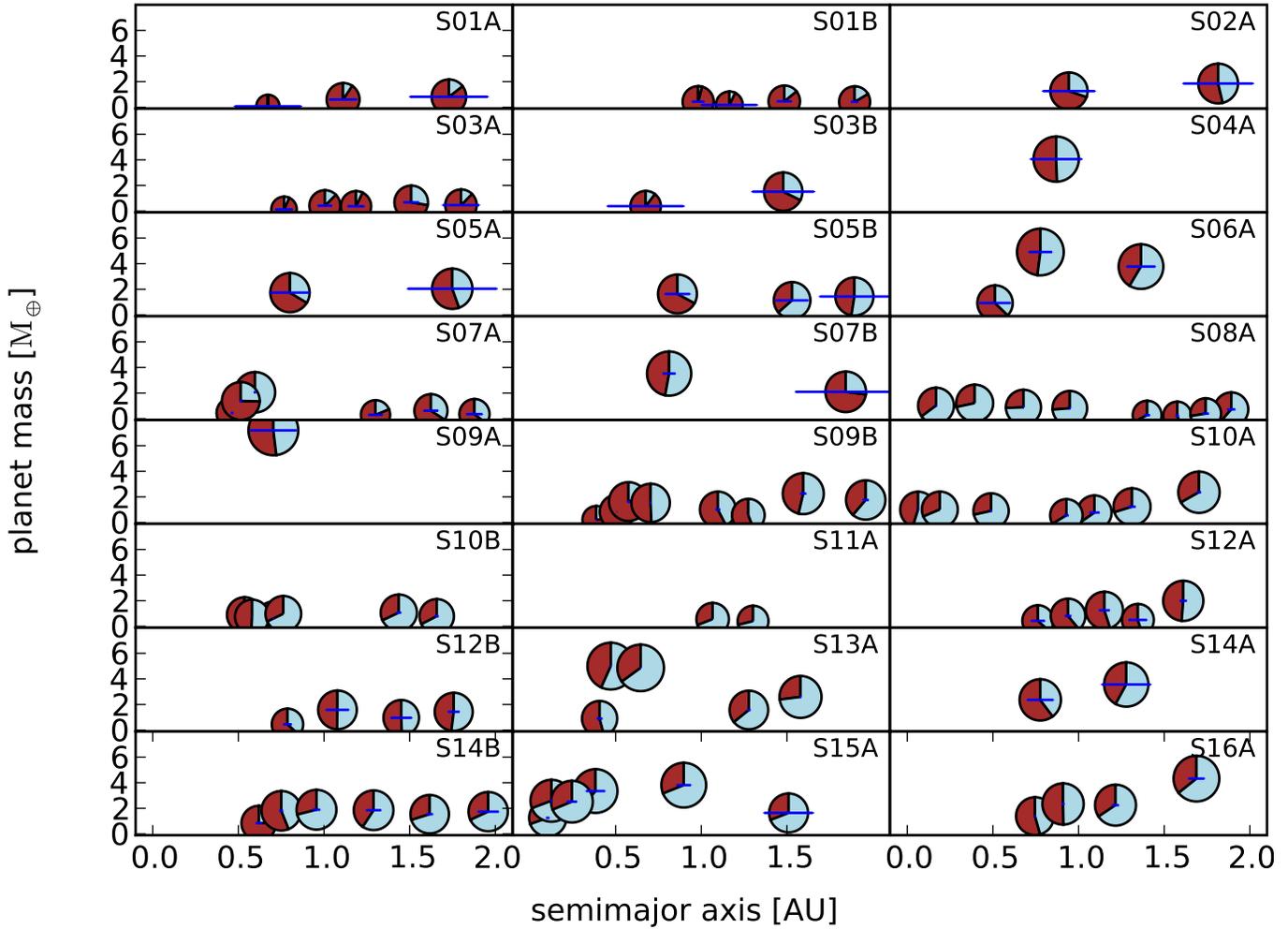}
      \caption{Final planetary systems at t = 100 Myr by simulation,
        planet mass in Earth masses versus semimajor axis in AU.  Each
        planet is plotted as a pie with radius proportional to its
        physical radius, with the blue (brown) slice corresponding to
        the fraction of its mass in ice (rock).  Horizontal bars
        describe the eccentric excursion, i.e. they extend from $a
        (1-e)$ to $a (1+e)$.  The labels specify the simulation
        reference number as listed in table
        \ref{simparam}. \label{plot:all_ma0}}
      
    \end{center}
  \end{figure*}

  \begin{figure*}
    \begin{center}
      \includegraphics[width=180mm, clip]{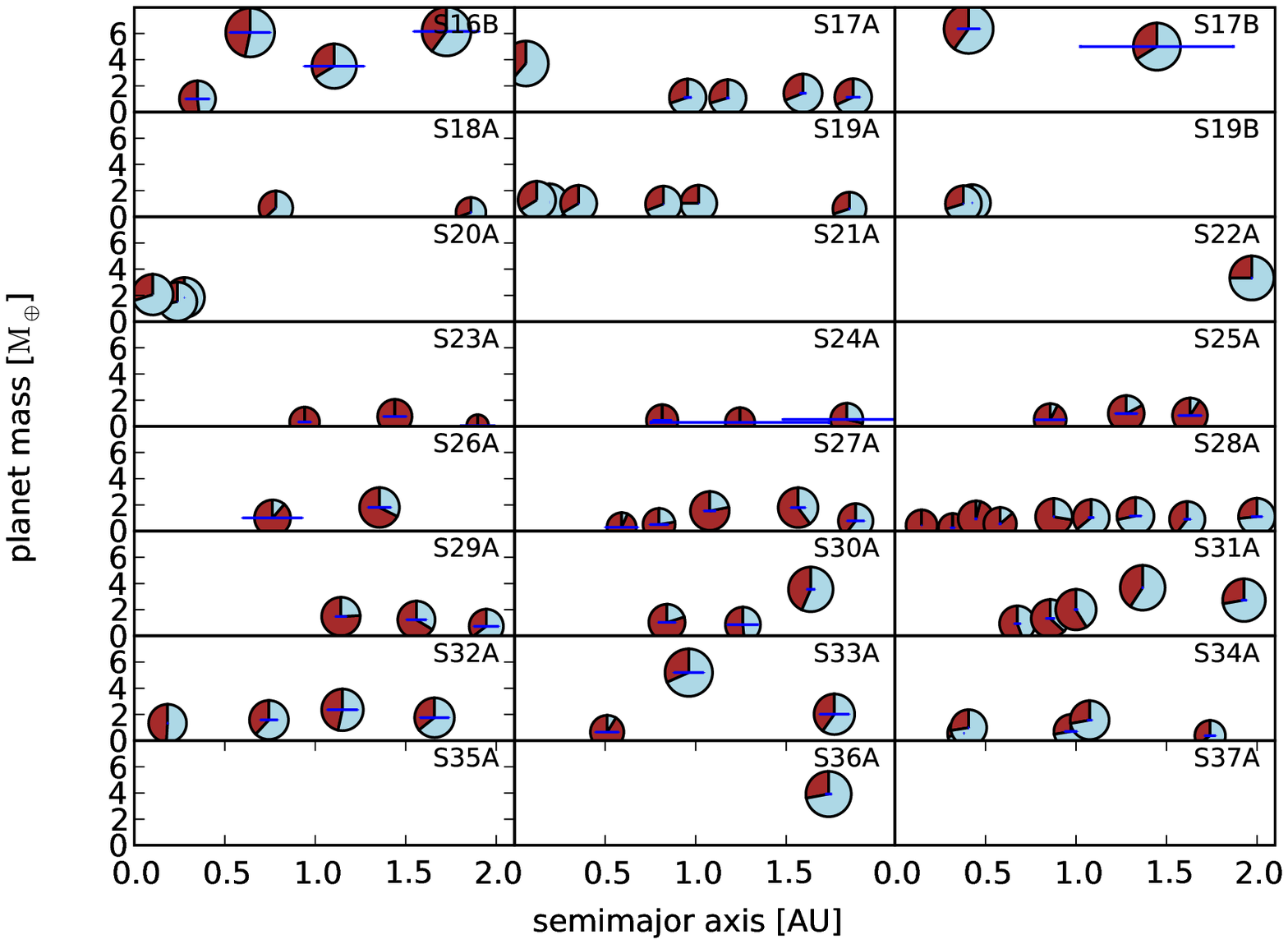}
      \caption{Final planets at t = 100 Myr; as in
        fig.~\ref{plot:all_ma0}.\label{plot:all_ma1} }
    \end{center}
  \end{figure*}

  Snapshots of the planetary systems which resulted at T=100 Myr are
  shown in figures \ref{plot:all_ma0} and \ref{plot:all_ma1}, with
  each planet being represented by a pie showing its fraction of rock
  and ice, and with horizontal bars indicating the eccentric excursion
  (and not, as is often done, $\pm5$ Hill radii.)  It is worth noting
  that the maximum ice fraction possible is 0.75.  Bodies composed
  entirely of material whose provenance lay interior to the snow line
  are 100\% rock.  Several general features of the simulation outcomes
  are immediately apparent in these figures.

  Most of the objects, especially the smallest ones, have modest
  eccentricity.  S07B and S17B are exceptions, in which the outermost
  body has a nonneglible $e$ (0.16 and 0.29, respectively.)  There is
  also the case of S24A, which is manifestly unstable.  The two outer
  planets had been exchanging $a$ and $e$ since an order-swapping
  encounter at 70 Myr, and had suffered a series of particularly close
  encounters $\simeq5$ Myr before the end of the simulation.  It might
  have been expected that all of the larger objects would have the
  smallest eccentricities due to dynamical friction and type I drag,
  but once the gas disc has dissipated and the local planetesimals
  have been consumed, the remaining large planets can generate
  significant eccentricities.

  It is also clear that stochasticity is playing a very important
  role.  While in some cases the same initial conditions generate
  similar final configurations even after the variation of the angles
  (e.g.~S01A/B, S05A/B), in other cases the resulting systems have
  almost no resemblance to each other (e.g.~S09A/B, S14A/B, S17A/B).
  This confirms the regrettable fact that the simulation outcomes are
  sufficiently sensitive to the details of the encounter and merger
  history that isolated runs convey limited information, although they
  can certainly demonstrate that a given scenario is possible.

  \begin{figure}
     \begin{center}
     \caption{Set of final planets at t = 100 Myr, planet mass versus
      semimajor axis.  Pies as in fig.~\ref{plot:all_ma0}.
      \label{plot:all_rockice}
      }
     \includegraphics[width=90mm, clip]{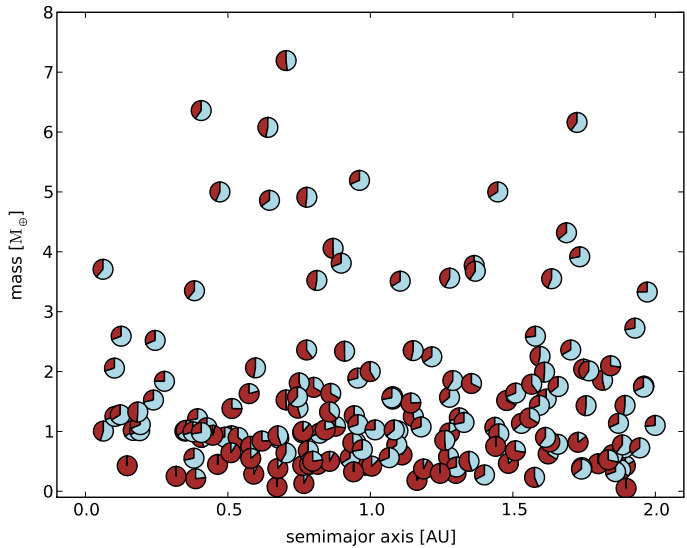}
     \end{center}
  \end{figure}

  Figure \ref{plot:all_rockice} shows the complete collection of
  resulting planets (combined from all simulations).  The majority of
  objects we form are small: the median planet mass is $1.07$ \ME, and
  90\% of the objects have masses below $3.55$ \ME.  Defining success
  as the production of an object with greater than 4 \ME interior to 2
  AU, there were only eight successful runs: S04A, S06A, S09A, S13A (2
  objects), S16A, S16B (2), S17B (2), and S33A.  Only one object (in
  S33A) came from a run with $S_{\mathrm{loc}}$ = 4 AU, and so our
  early, mildly encouraging experiments with a more distant snow line
  proved unfruitful in the production runs.  5 of the 7 successes (not
  double-counting S16) had $\fenh=5$, 2 had $\fenh$ = 3, and --
  significantly -- none had $\fenh = 10$.  5 of the 7 successes had
  $c_a = 0.3$.  Every $\alpha$ value generated at least one success
  (2, 3, 2 for $\alpha = 0.001, 0.5, 1.0$, respectively), and likewise
  for $\taudecay$ (4 for 1 Myr, 3 for 2 Myr).  The single most
  successful parameter set was that associated with S16A and S16B,
  namely $\fenh = 5, \alpha = 0.5, c_a = 0.3, \taudecay$ = 1 Myr,
  $S_{\mathrm{loc}}$ = 2.7 AU.

  \begin{figure}
    \begin{center}
    \caption{Summary of planet mass outcomes.
      \label{plot:mass_hist}
    }
    \includegraphics[width=90mm, clip]{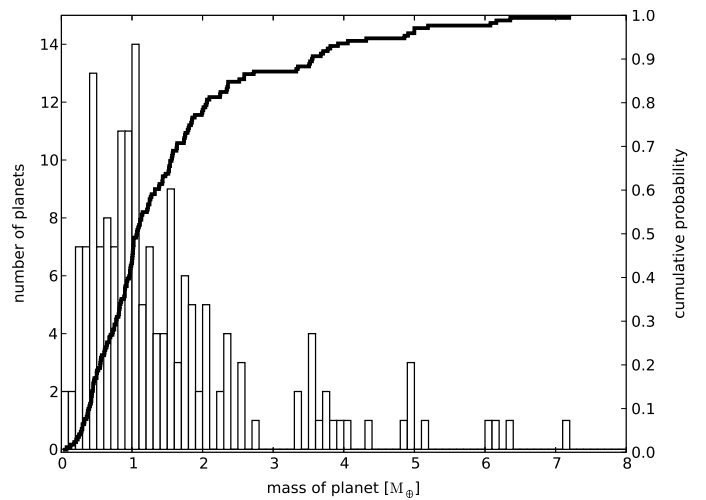}
    \end{center}
  \end{figure}
  
  It bears noting that the combined plots and histograms that we
  present in this section should be interpreted as `population
  synthesis' only in a loose sense.  There have been no corrections
  made for the shape of the parameter space, which includes
  duplications of some parameter runs, and which was not chosen to
  match any expected distribution of disc parameters in actual
  protoplanetary systems.  Accordingly, they should be taken merely as
  summaries of our particular outcomes, and not as predictions for
  what any real-world observer would see.
  
  \subsubsection{Planetary mass distribution}
  \label{section:mass-dist}

  Keeping that warning in mind, figure \ref{plot:mass_hist} shows the
  number of bodies of a given mass produced across all the
  simulations.  As mentioned previously, half of the planets formed
  have mass $\leq1$\ME.  Beyond $3\,$\ME, the numbers fall off
  dramatically.  This is in agreement with some (unpublished) early
  exploratory low-resolution runs which had great difficulty forming
  planets greater than $4$\ME regardless of the disc enhancement used.
  The median mass reached in the simulations is a full
  order of magnitude -- 17 times -- smaller than Neptune.

  We can also consider not only the successful runs (in which we have
  small-number statistics problems) but look at the complete dataset:
  see table \ref{medians}.  Perhaps unsurprisingly, the greater the
  enhancement factor, the greater the median mass.  However, this
  increase in the median value does not correspond to an increase in
  the maximum; indeed, the greater the enhancement $\fenh$, the {\em
    smaller} the resulting maximum mass of a planet.  A more robust
  measure is the change in the mean mass of the top quintile, which
  shows that $\fenh=5$ produced more large bodies than $\fenh=10$.  A
  similar wrap-around occurs with $\alpha$, where $\alpha=0.5$ has a
  higher median and top-quintile-mean than either $\alpha = 0.001$ or
  $\alpha = 1.0$.  The existence of such maxima in parameter space
  suggests that some natural methods for increasing the planet growth
  rate -- such as simply increasing the initial surface density of the
  disc -- may not help increase the final mass, as there are feedback
  mechanisms operating which resist the formation of larger bodies
  (namely the rapid inward migration of massive bodies which form
  early in a gas-rich environment).  In part, this result could merely
  be statistical noise, as of the 7 simulations with $\fenh=10$, only
  3 produced planets (and all of those had weak migration with $c_a =
  0.3$), but this low number is itself an example of the problem.
  This agrees with the predictions of \cite{komII}, and (except at
  very high disc masses) with the results of \cite{cham08}.
  
  Table \ref{medians} reveals several other (apparent) correlations
  between the simulation parameters and the resulting planet masses
  beyond those involving $\fenh$ and $\alpha$.  Decreasing the
  migration efficiency ($c_a$) improved the final mass, changing the
  median mass from $1.0 \ME$ to $1.4 \ME$ and the top quintile mean
  from $3.0 \ME$ to $4.5 \ME$.  The use of a more distant
  $S_{\mathrm{loc}}=4$, contrary to some of our early experiments,
  tended to reduce the planet mass.  Finally, the runs with
  $\taudecay$ = 2 Myr had smaller masses than those with $\taudecay$ =
  1 Myr, suggesting that the ability to bring in mass from farther out
  in the disc can overcome the loss of material to the star (or, at
  least, the simulation edge.)  However, given the wide variation in
  outcomes for the same parameter set -- and our use of no more than
  two instantiations of each -- these results should be treated
  cautiously, despite our averaging over the suite.

  \begin{table}

    \caption{Summary of resulting planets by parameter; values are
      taken over all planets produced in the subset of runs with
      parameter = value. \label{medians}}
    \begin{tabular}{@{}lllll@{}}
      \hline 
      Parameter & Value & \multicolumn{3}{c}{planetary mass statistics [\ME]} \\
       &  & Median & Maximum & Mean of top quintile\\
      \hline
      $f_{\mathrm{enh}}$ & 3  & 0.826& 7.194& 2.667 \\
      $f_{\mathrm{enh}}$ & 5 & 1.455& 6.359& 4.428 \\
      $f_{\mathrm{enh}}$ & 10 & 2.058& 3.917& 3.917 \\
      $\alpha$ & 0.001  & 1.025& 5.000& 3.353 \\
      $\alpha$ & 0.5  & 1.412& 6.359& 4.478 \\
      $\alpha$ & 1.0  & 0.987& 7.194& 2.878 \\
      $S_{\mathrm{loc}}$ & 2.7 AU  & 1.124& 7.194& 3.980 \\
      $S_{\mathrm{loc}}$ & 4.0 AU  & 0.982& 5.193& 2.901 \\
      $c_a$ & 0.3  & 1.430& 7.194& 4.482 \\
      $c_a$ & 1.0  & 1.014& 6.359& 3.030 \\
      $\taudecay$ & 1 Myr  & 1.044& 7.194& 3.447 \\
      $\taudecay$ & 2 Myr& 1.250& 5.000& 4.402 \\
    \end{tabular}

  \end{table}

  In some runs, a considerable amount of mass was removed from the
  simulation by falling off our inner boundary of 0.05 AU.  In the
  \fenh=10, $S_{\mathrm{loc}}$=2.7 AU runs S20A, S21A, and S22A, the
  lost planets totalled 34 \ME (with the maximum mass of a lost planet
  being 5.2 \ME), 38 \ME (maximum 4.0 \ME), and 67 \ME (maximum 7.1
  \ME).  (The continuous, fixed-spacing semianalytics generating the
  initial conditions are probably unreliable at such a large
  enhancement.)  In the $\fenh=3$ and $\fenh=5$ cases, the amount of
  mass lost varied from none (in all the discs with $\alpha=0.001$
  except S15A, which had $c_a=1$ and \taudecay = 2 Myr) to 21-24 \ME
  in S18A and S19A/B.  Excluding the \fenh=10 cases, none of the lost
  bodies were larger than 3.3 \ME (in S16A), and the median mass of a
  body which was lost was 1.9 \ME.
  
  \subsubsection{Ice fractions}
  \label{section:ice-fraction}

\begin{figure}

    \caption{Summary of ice fraction outcomes.
        \label{plot:rock_hist}
    } \includegraphics[width=90mm, clip]{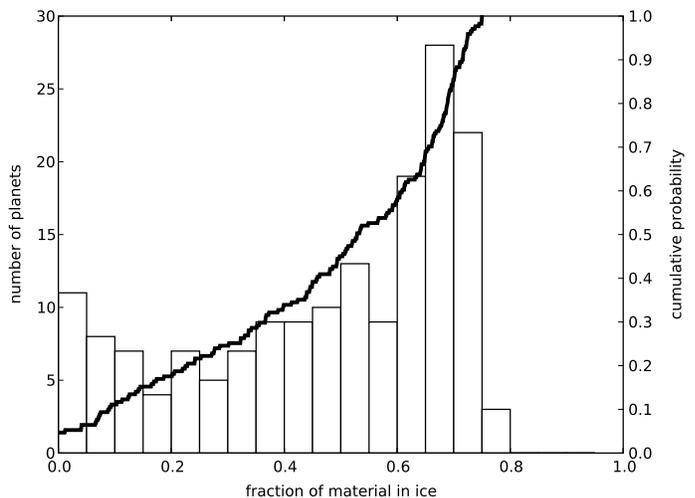}

  \end{figure}

  Figure \ref{plot:all_rockice} demonstrates that larger-mass objects
  tend to have higher ice fractions, and given our initial assumption
  that 75\% of the material past the snow line was in ices, the ice
  fraction is a proxy for the radial transport.  The median ice
  fraction for objects greater than 1 \ME$\,$ is 0.60, and for objects
  smaller, 0.36.  There were only 7 objects which contained no ices at
  all (1 from S01A; 3 from S23A; 2 from S24A; 2 from S28A), and all
  but one of those had masses smaller than 0.5 \ME, the exception
  having 0.75 \ME.  Every ice-free body was formed from a run with
  $\fenh = 3$ and $\taudecay$ = 1 Myr, suggesting that we should only
  expect to find completely rocky bodies in low-enhancement discs with
  short disc lifetimes.  Moreover, all but one ice-free planet was
  from a simulation with $S_{\mathrm{loc}}$ = 4 AU, which suggests
  that in the more physically realistic scenarios we should expect
  almost no bodies which fail to contain material from beyond the snow
  line.  (Of course, `ice-free' and `completely rocky' here are
  relative to the assignment of ice fractions at t=0.4 Myr, when our
  N-body integrations began, although they should be assigned in a way
  consistent with the snow enhancement.)  It is notable, as shown in
  figure \ref{plot:rock_hist}, that the entire range of possible ice
  fractions is covered, from bodies which are completely ice-free to
  bodies which reach the maximal 75\% value.

\begin{figure}

    \caption{Ice fraction versus amount of material consumed by an
      object in the form of an embryo (as opposed to a planetesimal).
      By construction all objects have ice fractions $\leq$ 0.75.
      \label{plot:emb_vs_rock}}
    \includegraphics[width=90mm, clip]{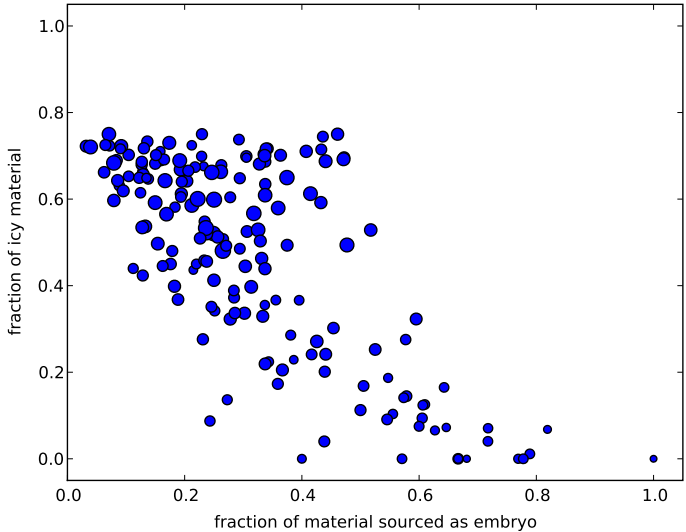}
  \end{figure}

  Figure \ref{plot:emb_vs_rock} shows that the lower the ice fraction,
  the more of an object's mass was consumed (or started) as an embryo,
  and not a planetesimal.  At the start of the simulation, all embryos
  are either inside the snow line, in which case we assume they are
  composed entirely of rock, or outside the snow line, in which case
  we assume they are composed of 75\% ice and 25\% rock.  (We neglect
  the early semianalytic evolution of the models before t = 0.4 Myr;
  as a result, we slightly underestimate both the initial ice
  fractions of embryos in the inner regions and their consumption of
  planetesimals.)  Accordingly, at t=0.4 Myr, all embryos are either
  located at (1, 0) or (1, 0.75) on this plot: they are composed
  entirely of embryo material, and have ice fractions of either 0 or
  0.75.

  At first, objects located in both regions accrete local material
  from both embryos and planetesimals, which decreases their embryo
  fractions, but leaves their ice fractions constant as they accrete
  from nearby material which has the same ice fractions as they do
  (whether 0 or 0.75).  Both interior and exterior objects therefore
  move left on the diagram.  Eventually migration becomes important,
  which tends to move ice-rich material into the inner regions.
  Accretion then raises the ice fractions of the rocky bodies (which
  now have icy material to consume) and lowers the ice fractions of
  the icy bodies (which previously had mostly icy material available
  but now find more rocky material in their feeding zone).  Since so
  much material is brought into the 0.05--2 AU zone from outside, this
  tends to concentrate the resulting planets at high ice fraction with
  roughly 20--40\% of their mass coming from embryos (the median
  embryo mass fraction is 0.28).  Almost all objects with very low
  embryo mass fraction ($<0.2$), which therefore consumed almost all
  of their material in planetesimals, have very high ice fraction.
  These values may be biased somewhat by our choice of inner edge,
  which artificially depletes the innermost regions of rocky
  planetesimal material.

  In the innermost regions, $a < 0.25$ AU, somewhat
  counter-intuitively, we find smaller bodies with higher ice content.
  All objects save one (an 0.43 \ME ice-free object from S28A) had
  masses 1 \ME $< M < 4$ \ME, and were composed of at least 50\% ice.
  Excluding the rocky outlier, the median ice fraction was 0.69.
  Since the maximum ice fraction possible in our simulations is 0.75,
  these objects are composed almost entirely of material from beyond
  the snow line.

  \subsection{Simulation statistics}
  \label{section:chambstat}
  \cite{cham01} (\S 4, which we follow closely here) introduced a
  useful set of dimensionless statistics for comparing the results of
  terrestrial planet simulations, but as he notes they are of general
  applicability.  (Given our inability to build planets larger than 8
  \ME, the statistics are rather more closely applicable than we had
  hoped.)  His set includes:
  \begin{enumerate}
    \item $N$, the number of bodies (here planets).
    \item $S_m$, the fraction of the total mass in the largest object.
    \item $S_s$, a spacing statistic, loosely related to the empirical
      instability time-scale (\citealt*{stab1}; see also
      \citealt*{stab2} for a more recent investigation of the problem,
      which suggests that the dependence on the mass should be closer
      to $0.29 \simeq 2/7$), defined by
      \begin{equation}
        S_s = \frac{6}{N-1} \left( \frac{a_{\mathrm{max}} - a_{\mathrm{min}}}{a_{\mathrm{max}} + a_{\mathrm{min}}} \right)
        \left( \frac{3 M_\Sun}{2 \bar{m}} \right)^{1/4}
      \end{equation}
      where $a_{\mathrm{min}}$ and $a_{\mathrm{max}}$ are the limiting
      semimajor axes (here set to 0.05 AU and 2.0 AU), and $\bar{m}$
      is the mean planet mass.
    \item $S_d$, the normalized angular momentum deficit, defined by
      \begin{equation}
        S_d = \frac{\Sigma m_j \sqrt{a_j} ( 1 - \sqrt{1 - e_j^2} \cos i_j )}
        {\Sigma m_j \sqrt{a_j}}
      \end{equation}
      with summation over the planets, indexed by $j$. (See, e.g., \citealt{laskar1997}.)
    \item $S_c$, a concentration statistic, given by
      \begin{equation}
        S_c = \max \left( \frac{ \Sigma m_j }{ \Sigma m_j \left[
            \log_{10} (a/a_j) \right]^2 } \right)
      \end{equation}
      where we take the maximum value the argument reaches over all
      values of the variable $a$ in the interval [0.05 AU, 2 AU].  The
      higher the value, the more `concentrated' the radial
      distribution is, although this statistic is not a proxy for the
      mass surface density.
    \item $S_r$, a radial mixing statistic, given by
      \begin{equation}
        \label{eqn:S_r}
        S_r = \frac{1}{\Sigma m_j} \sum \frac{m_j |a_{j,0} - a_{j, f}|}{a_{j, f}}
      \end{equation}
      where $0$ and $f$ refer to the original (at t=0.4 Myr) and final
      (at t=100 Myr) semimajor axes of each object which ultimately
      becomes part of one of the resulting planets.  This statistic
      was introduced in \cite{cham01} to quantify the degree of mixing
      induced by gravitational scattering.  Being linear in the change
      in semimajor axis, this statistic may not prove as useful
      measuring this effect in simulations with high migration rates,
      but will quantify the degree of type I migration instead.
  \end{enumerate}

  These statistics, along with the total mass $\Mtot$ and mean mass
  $\bar{m}$, are listed in table \ref{chambtable}, as well as the
  corresponding values for our terrestrial system (Mercury, Venus,
  Earth, and Mars) and the system of three Neptune analogues orbiting
  HD69830.

  \begin{table}

    \caption{Summary statistics for each run; definitions as in \S
      \ref{section:chambstat}.  For comparison, the label ``TERR''
      describes the four terrestrial planets in our own system, and
      3NEP the three Neptunes in the HD69830 system.  Undefined values
      are labelled N/A, and unknown values left blank. Simulations
      S21A, S35A, and S37A resulted in no planets interior to 2 AU,
      and so are not listed.  \label{chambtable}}

     {\tiny
     \begin{tabular}{p{0.3cm}p{0.24cm}rrrrrrr}
       \hline 
       Sim. & N & \hspace{-0.5cm} $\Mtot$ & $\bar{m}$ & $S_m$ & $S_s$ & $S_d$ & $S_c$ & $S_r$  \vspace{0.05cm}\tn

       \hline
       S01A & 3 & 1.50 & 0.50 & 0.55 & 90.23 & 0.008 & 73.55 & 0.44 \tn
       S01B & 4 & 1.51 & 0.38 & 0.31 & 64.54 & 0.003 & 79.20 & 0.46 \tn
       S02A & 2 & 3.12 & 1.56 & 0.59 & 135.79 & 0.014 & 51.42 & 1.35 \tn
       S03A & 5 & 2.06 & 0.41 & 0.33 & 47.35 & 0.002 & 84.02 & 0.59 \tn
       S03B & 2 & 1.90 & 0.95 & 0.80 & 153.74 & 0.018 & 53.78 & 0.97 \tn
       S04A & 1 & 4.06 & 4.06 & 1.00 & N/A & 0.016 & N/A & 2.54 \tn
       S05A & 2 & 3.78 & 1.89 & 0.54 & 129.43 & 0.022 & 35.04 & 1.49 \tn
       S05B & 3 & 4.20 & 1.40 & 0.39 & 69.73 & 0.005 & 43.67 & 1.51 \tn
       S06A & 3 & 9.61 & 3.20 & 0.51 & 56.71 & 0.007 & 48.89 & 3.05 \tn
       S07A & 6 & 5.17 & 0.86 & 0.40 & 31.49 & 2.1e-04 & 23.62 & 2.99 \tn
       S07B & 2 & 5.62 & 2.81 & 0.63 & 117.20 & 0.014 & 33.75 & 2.31 \tn
       S08A & 8 & 5.56 & 0.69 & 0.22 & 23.74 & 4.5e-05 & 7.67 & 10.02 \tn
       S09A & 1 & 7.19 & 7.19 & 1.00 & N/A & 0.023 & N/A & 3.49 \tn
       S09B & 8 & 9.78 & 1.22 & 0.23 & 20.61 & 1.4e-04 & 20.13 & 2.10 \tn
       S10A & 7 & 7.84 & 1.12 & 0.30 & 24.58 & 1.1e-04 & 4.10 & 10.65 \tn
       S10B & 6 & 5.12 & 0.85 & 0.21 & 31.57 & 2.6e-05 & 27.84 & 3.07 \tn
       S11A & 2 & 0.97 & 0.48 & 0.58 & 181.97 & 9.4e-07 & 544.68 & 3.44 \tn
       S12A & 5 & 4.98 & 1.00 & 0.40 & 37.97 & 3.8e-04 & 84.87 & 1.94 \tn
       S12B & 4 & 4.42 & 1.10 & 0.36 & 49.33 & 0.001 & 75.57 & 1.83 \tn
       S13A & 5 & 14.92 & 2.98 & 0.34 & 28.86 & 9.4e-05 & 23.58 & 5.62 \tn
       S14A & 2 & 5.92 & 2.96 & 0.60 & 115.67 & 0.007 & 88.91 & 2.82 \tn
       S14B & 6 & 9.67 & 1.61 & 0.20 & 26.93 & 6.2e-04 & 36.31 & 2.73 \tn
       S15A & 6 & 15.16 & 2.53 & 0.25 & 24.07 & 0.007 & 6.86 & 16.94 \tn
       S16A & 4 & 10.29 & 2.57 & 0.42 & 39.93 & 2.7e-04 & 54.90 & 2.53 \tn
       S16B & 4 & 16.75 & 4.19 & 0.37 & 35.35 & 0.012 & 21.20 & 3.39 \tn
       S17A & 5 & 8.47 & 1.69 & 0.44 & 33.25 & 2.1e-04 & 2.21 & 30.17 \tn
       S17B & 2 & 11.36 & 5.68 & 0.56 & 98.28 & 0.041 & 13.34 & 6.12 \tn
       S18A & 2 & 1.01 & 0.50 & 0.67 & 180.19 & 5.0e-07 & 32.16 & 3.74 \tn
       S19A & 6 & 6.02 & 1.00 & 0.21 & 30.32 & 1.0e-06 & 6.10 & 14.58 \tn
       S19B & 2 & 2.05 & 1.03 & 0.52 & 150.72 & 9.3e-06 & 1426.51 & 11.62 \tn
       S20A & 3 & 5.42 & 1.81 & 0.38 & 65.44 & 2.7e-06 & 25.43 & 33.28 \tn
       S22A & 1 & 3.33 & 3.33 & 1.00 & N/A & 2.4e-06 & N/A & 2.05 \tn
       S23A & 3 & 1.12 & 0.37 & 0.67 & 96.94 & 0.002 & 122.19 & 0.23 \tn
       S24A & 3 & 1.27 & 0.42 & 0.41 & 93.96 & 0.034 & 42.33 & 0.45 \tn
       S25A & 3 & 2.30 & 0.77 & 0.42 & 81.07 & 0.002 & 94.80 & 0.67 \tn
       S26A & 2 & 2.80 & 1.40 & 0.64 & 139.49 & 0.011 & 70.28 & 1.32 \tn
       S27A & 5 & 4.89 & 0.98 & 0.37 & 38.14 & 0.001 & 50.19 & 1.28 \tn
       S28A & 9 & 7.42 & 0.82 & 0.15 & 19.90 & 2.1e-04 & 11.39 & 2.70 \tn
       S29A & 3 & 3.42 & 1.14 & 0.43 & 73.39 & 6.3e-04 & 119.09 & 1.36 \tn
       S30A & 3 & 5.42 & 1.81 & 0.65 & 65.42 & 9.7e-04 & 79.77 & 1.89 \tn
       S31A & 5 & 10.65 & 2.13 & 0.35 & 31.40 & 1.4e-04 & 47.78 & 2.35 \tn
       S32A & 4 & 7.00 & 1.75 & 0.34 & 43.97 & 0.002 & 9.02 & 7.42 \tn
       S33A & 3 & 7.85 & 2.62 & 0.66 & 59.65 & 0.007 & 45.37 & 3.39 \tn
       S34A & 5 & 4.16 & 0.83 & 0.38 & 39.71 & 2.2e-04 & 19.50 & 6.39 \tn
       S36A & 1 & 3.92 & 3.92 & 1.00 & N/A & 6.6e-05 & N/A & 2.29 \tn
       \hline
       TERR & 4 & 1.98 & 0.49 & 0.51 & 6.29 & 0.002 & 89.49 & \tn
       3NEP & 3 &  41.00& 13.67 & 0.45& 4.12& 0.004 & 7.05 & \tn
       \end{tabular}
     }

  \end{table}

  \subsubsection{Final number of planets and radial mixing}
  \label{section:N}

  A somewhat surprising outcome of the simulations is that there were
  so many systems with high N which survived: 9 of the 48 simulations
  resulted in systems with N $>$ 5 even after 100 Myr. Some of this
  surprise is explained by the fact that intuitions trained on the
  terrestrial regime can fail for $a < 0.5$ AU, where the same radial
  separation corresponds to a much greater Hill separation, and many
  of the high-N systems have multiple objects in the interior.
  Removing these objects from consideration reduces the number of
  high-N systems to 5.  That said, these systems (both the original 9
  and the reduced 5) cover every $\alpha$, include $\fenh = 3$ and 5,
  both $\taudecay$ values, both $S_{\mathrm{loc}}$ values, and all but
  one have $c_a=1.0$.  This ubiquity suggests that the unexpectedly
  large number of planets is likely to be the result of strong
  migration selecting stable configurations.

\begin{figure}
    \caption{Number of planets versus radial mixing parameter.\label{fig:N_vs_S_r}}
    \begin{center}
      \includegraphics[width=90mm, clip]{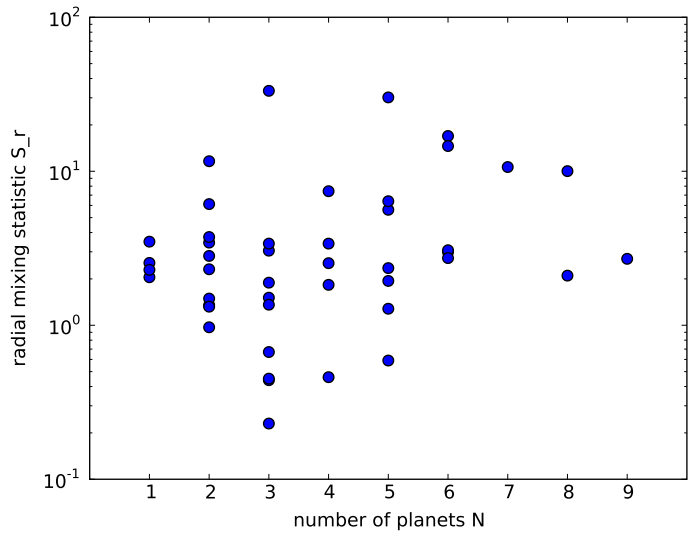}
    \end{center}
  \end{figure}

   When applied to our runs, the radial mixing parameter $S_r$ defined
   by equation~\ref{eqn:S_r} provides a measure of the degree of
   migration experienced by all components of the final planetary
   system. For example, a (somewhat unrealistic) system of two equal
   mass planets which begin life with semimajor axes $a \simeq 2$ AU,
   and subsequently migrate inward without further accretion to the
   radial distance $\simeq 0.2$ AU will have a value $S_r \simeq 10$
   (this value becomes $S_r \simeq 20$ if the final stopping location
   decreases to 0.1 AU; it becomes $S_r=1$ if the final stopping
   distance increases to 1 AU).  As figure \ref{fig:N_vs_S_r} shows,
   while there are many low-N runs with $S_r$ comparable to the high-N
   values, there are no high-N runs with low $S_r$.  Of the 9 N $>$ 5
   runs, all had radial mixing $S_r > 2$, with 4 having $S_r > 10$
   (44\%); of the 40 $N \le 5$ runs, 40\% had $S_r < 2$, and only 3
   had $S_r > 10$ (7\%).  This suggests -- albeit weakly, given the
   small numbers involved -- that high-N systems consist of bodies
   that formed over a wide range of radii and migrated into the
   interior region.  There are also no high-N runs with $\Mtot < 5$
   \ME (numerous planets add up to a significant amount of total
   mass). The absence of any high-N systems with $\fenh = 10$ is
   likely the result of the would-be planets migrating out of our
   integration region, and another example of how in this regime
   higher surface densities can hurt more than they help.  Indeed, all
   three runs which resulted in no planets interior to 2 AU (S21A,
   S35A, and S37A) all had $\fenh = 10$ and $c_a = 1$, although here
   our choice of outer boundary may be playing a role; we return to
   this issue in \S\ref{weaknesses}.
   
  It is probable that the perturbations from giant planets would
  significantly lower these numbers.  Their formation is not modelled
  here: accretion of gas is not treated, nor are any objects of $a >$
  2 AU after 6 Myr. \cite{cham01} notes that one explanation
  consistent with the observed decrease in terrestrial planet number
  between his simulations and those of some previous authors was his
  incorporation of the Jovians, and a similar effect should occur
  here.  \cite{mcn05} also had to invoke an external random velocity
  source to reduce the number of planets in late-stage
  prototerrestrial systems produced via oligarchic migration.  Of
  course, even had we incorporated a gas accretion model, we do not
  form any objects large enough to serve as useful seeds for the
  \cite{poll} picture while the gas is present.

  \subsubsection{Total surviving mass}
  \label{section:surviving_mass}

   Of primary importance for our purposes is the total amount of mass
   which survives interior to 2 AU at t=100 Myr.  Only 6 runs
   succeeded in producing more than 10 \ME of material inside 2 AU.
   Despite the small number, it is striking that every one had
   $\fenh=5$ and $S_{\mathrm{loc}} = 2.7$ AU, with $\alpha < 1.0$.
   S16 -- with $\alpha = 0.5, c_a = 0.3, \taudecay=1$ Myr -- was
   particularly successful, with both instantiations producing over 10
   \ME, and S16B producing two of the largest bodies in the suite and
   the single largest total mass, 16.75 \ME.  Recall that for our
   vertical profile (eq.~\ref{eq:z0}), $\alpha = 1.0$ is the
   transition between convergent and divergent migration for
   equal-mass bodies, and that for $\alpha < 1.0$, neighbouring
   equal-mass migrating objects converge.  At the outset of this
   project we had hoped that the compression might increase the
   resulting masses, and both the behaviour at the high $\Mtot$ limit
   and summing over the other parameters as in table \ref{medians}
   support this hope.  When the $\Mtot$ limit is lowered to 5 \ME, 7
   of the 23 runs (30\%) had $\alpha=1.0$, and so any advantage that
   convergent values of $\alpha$ possess seems limited to the extremes
   (which is where one might expect to first observe increases in the
   variance of the mass with a lowered $\alpha$.)

  \begin{figure}
    \caption{Number of planets versus total and mean planet
      mass.\label{fig:N_vs_Mtot}}
    \begin{center}
      \includegraphics[width=90mm, clip]{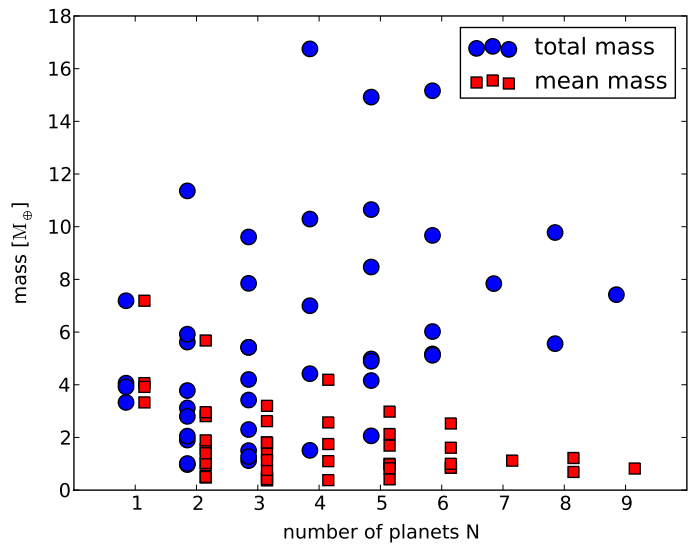}
    \end{center}
  \end{figure}

  Figure \ref{fig:N_vs_Mtot} shows the variation of $\Mtot$ and
  $\bar{m}$ with the number of planets.  The cluster of high-mass runs
  with $\Mtot \simgeq$15 \ME had nearly the median number of planets
  ($4 \leq N \leq 6$), while both lower and higher values of $N$ had
  lower maximum $\Mtot$ (11.4 \ME and 9.8 \ME, respectively).  This
  hints that median $N$ values produce the largest masses, but there are
  only three high-mass runs involved.  Moreover, the four runs with $N
  > 6$ also have mean and median $\Mtot$ greater than or comparable to
  the median runs, even if the maximum is greatly reduced
  (unsurprisingly, having more planets tends to translate into a
  greater total mass). It does appear that both the maximum value and
  the variance decrease with small and large $N$; there are no runs
  with $N \geq 6$ and $\Mtot <$ 5 \ME.  There is also a strong
  correlation between the number of resulting planets and their mean
  mass, such that the larger $N$ is, the lower $\bar{m}$ is.  The
  curves which bound $\bar{m}$ as $N$ varies are quite regular, which
  given the volume of parameter space that our search covered is
  strong evidence that we must make large modifications to our initial
  conditions or our physics to increase $\bar{m}$ to Neptune-like
  masses at $N \geq 3$.\footnote{It is darkly amusing to note that
    naively extrapolating the best fit line for the maxima predicts
    that the desired $\bar{m} \simeq 15$\ME occurs at $N\simeq-11$, a
    situation which should defeat even the admirable ingenuity of
    today's observational exoplanetary community.}

  \begin{figure}
    \caption{Total mass versus mean mass.\label{fig:Mtot_vs_Mmean}}
    \begin{center}
      \includegraphics[width=90mm, clip]{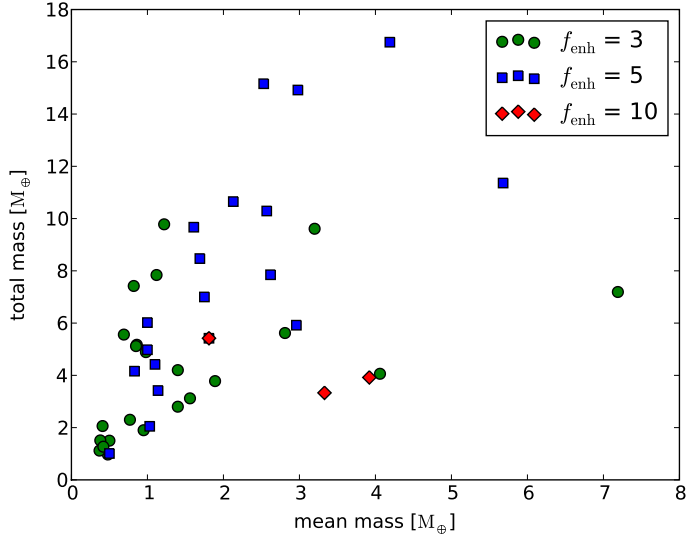}
    \end{center}
  \end{figure}
 
  One way to see the difficulty is to compare the total mass in a run
  with the mean mass (fig.~\ref{fig:Mtot_vs_Mmean}; zero-mass outcomes
  are suppressed.).  There is a cluster of very low-mass runs with
  $\bar{m}\sim\!0.5$\ME, but at larger $\bar{m}$ and $\Mtot$ the
  region of achieved values opens up.  Clearly $\bar{m} \leq \Mtot$,
  which sets the lower boundary curve.  Removing the high-$\bar{m}$
  outliers, the $\fenh=3$ and $\fenh=5$ cases show similar behaviour,
  except that the $\fenh=5$ set reaches higher total masses and has a
  larger variance.  Not only do the $\fenh = 10$ cases fail to improve
  on the $\fenh=5$ case, each is inferior to many $\fenh=3$ runs.
  Worse yet, three zero-mass runs were not plotted, although some of
  those had full migration and lost implausible amounts of mass to the
  star.  This leads us to tentatively conclude that any mechanism that
  increases the rate of planetary growth during early times is likely
  to reduce, and not increase, the mass of the largest surviving
  bodies in the presence of significant inward type I migration.

\begin{figure}
  \caption{Number of planets versus fraction in largest body.  The
  line corresponding to $1/N$ is plotted.
  \label{fig:N_vs_Sm}}
  \begin{center}
  \includegraphics[width=90mm, clip]{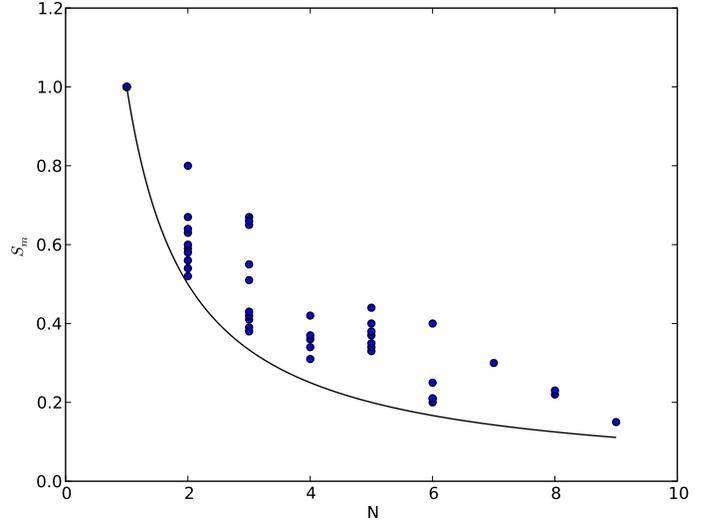}
  \end{center}
  \end{figure}

  Figure \ref{fig:N_vs_Sm} shows the fraction of the total mass
  contained in the largest body (by definition, this cannot be smaller
  than $1/N$).  Slightly under half of the planet-producing
  simulations have $S_m > 0.5$, and the trend toward lower $S_m$ with
  larger $N$ is clear.  For comparison, the current HD69830 value has
  $N=3, S_m=0.45$, well within the achieved range.

  \subsubsection{Future mergers and long-term stability}
  \label{section:mergers}
 
  \begin{figure}
    \caption{Mass and separation of neighbouring pairs at t=100
      Myr.\label{bseps}}
    \begin{center}
      \includegraphics[width=90mm, clip]{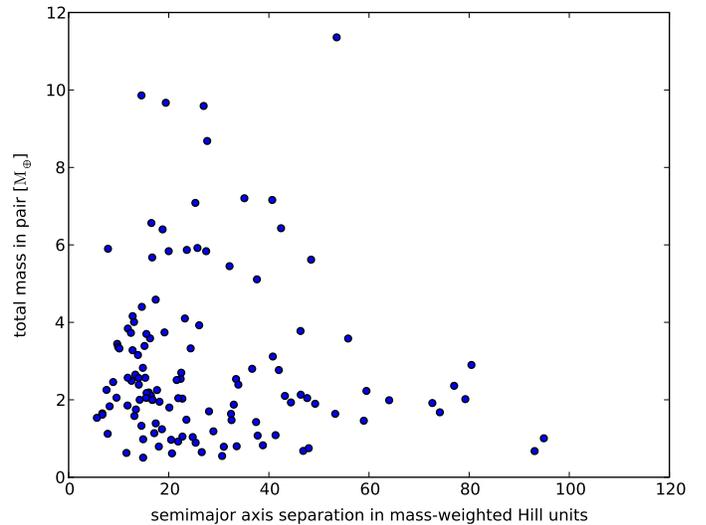}
    \end{center}
  \end{figure}
 
  \begin{figure}
    \caption{Evolution of median separation of planets interior to 2
      AU.  \label{plot:bvst}}
    \begin{center}
      \includegraphics[width=90mm, clip]{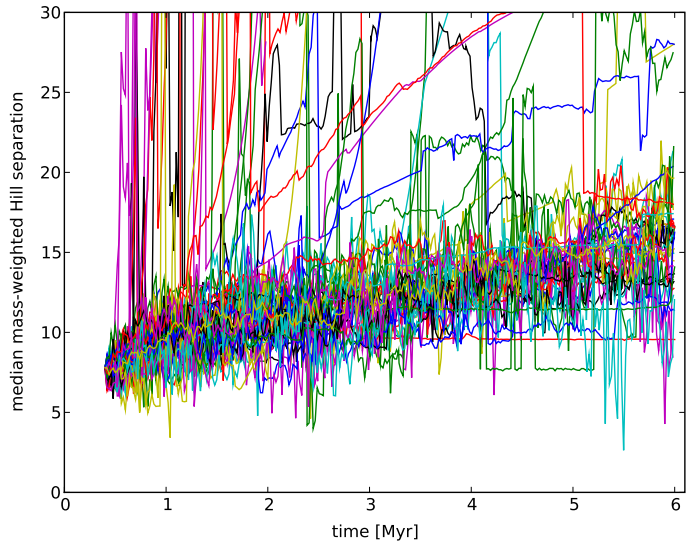}
    \end{center}
  \end{figure}
  
 It is unlikely that most of the resulting systems will undergo many
 more mergers (save S24A).  Figure \ref{bseps} shows the mass in each
 pair of neighbouring planets in all the resulting systems against the
 mass-weighted Hill separation ($\bar{b} = ((M_1 + M_2)/3
 M_{\Sun})^{1/3} (M_1 a_1 + M_2 a_2)/(M_1 + M_2)$) of the pair.  There
 are only 5 pairs of planets involving total masses greater than $8
 M_{\Earth}$, and 4 of those have $\bar{b} \gtrsim 20$.  The median
 $\bar{b}$ is $\simeq22$, and the median mass of objects closer than
 the median separation is only $2.3 M_{\Earth}$.  If we follow the
 mass-separation relationship for each pair through the simulations,
 we find that the vast majority of pairs which have small separations
 have combined masses $<$ 4 \ME (which helps explain why we do not
 form very many large objects).

 Figure \ref{plot:bvst} shows the evolution of the median
 mass-weighted Hill separation of all objects inside 2 AU for all
 simulations until 6 Myr.  While the data are quite noisy, it is clear
 that almost all simulations show a definite evolution from their
 initial value of $\simeq7$ ($\simeq9$ in single-planet units) to
 $\simeq 15$ ($\simeq 19$); the median outcome is 16.5.  33 of the 48
 simulations (69\%) have median $\bar{b} > 15$, and 14 have median
 $\bar{b} > 20$.  Only 6 of the runs (12.5\%) had median $\bar{b} <
 13$.  This difference can be significant as the interaction
 time-scale is a strong function of the separation.  Work by
 \cite{stab2} shows that the time-scale for a collection of cold
 equal-mass objects to suffer an instability in the absence of nebular
 gas when separated by 15-20 mutual Hill radii can be $>10^{10}$
 years.  These systems have much shorter interaction times in practice
 due to the variance of the spacing and the nonzero eccentricities
 \citep{zhou07}, but the likelihood of the set of planets merging to
 form Neptune-sized objects even when the total mass makes it possible
 is very low.  In any case, these total masses are far beneath the
 $\sim\!41 \ME$ present in the HD69830 system, and the maximum mass of
 $7.19\,$ \ME is far below the estimated 18.4 \ME of HD69830 d.

  \section{Discussion}
  \label{discussion}

  \subsection{Weaknesses of the model}
  \label{weaknesses}
  As usual, many of the approximations invoked to make the simulations
  tractable may have affected the results.

  The semianalytic model used to advance the early stages of the model
  is quite primitive.  It uses a fixed Hill spacing of 10 between the
  embryos, using neither the original \cite{ko98} dependence of $b$ on
  $M$ nor a more realistic slower growth.  This should be a tolerable
  error during the first 0.5 Myr for all but the highest-enhancement
  cases.  The approximation of the embryo distribution by a smooth
  function means that most of the interembryo dynamics is lost, which
  can be important when tidal convoys of migrating objects locked into
  mean-motion resonance form.  In a noisy N-body migration problem,
  objects which become larger than their neighbours due to a merger
  can push interior embryos with it at migration rates larger than the
  naive model would predict (as in S07A -- see figure~\ref{S07A_qaQ}).
  \cite{cham08} discusses this issue in the context of building
  semianalytic models of oligarchy.  The use of a large seed mass
  $M_0$ makes the discs easier to instantiate in the N-body code by
  lowering the gradient in embryo mass, but at the cost of
  artificially accelerating the growth of the most distant objects,
  which should lead to higher migration.  Quite substantial growth
  above this initial mass is required, however, to enable
  trans-snowline embryos to migrate interior to 2 AU, so this is
  unlikely to be a significant problem in the simulations.

  Enforcing a sharp embryo/planetesimal distinction such that
  planetesimals neither self-gravitate nor accrete, the usual
  procedure in the field among those using Kepler-problem symplectic
  integrators, results in a poor treatment of the mass spectrum.  No
  planetesimals can promote themselves to embryos, even if a
  planetesimal ring forms in which the largest body should experience
  runaway growth and become a new oligarch.  This will suppress embryo
  growth in some cases (when an oligarch should have formed), and
  increase it in others (by providing a fresh supply of material for
  protoplanets migrating into the region to consume, as in
  \citealt{growthofmig} and the models of \citealt*{alib}).  The use
  of a uniform characteristic planetesimal radius means that we are
  insensitive to the differing effects of aerodynamic drag on
  different-sized objects.  While the planetesimal non-accretion
  problem is difficult to overcome, the uniform-mass problem could be
  handled by varying the radius assigned to super-planetesimals, so
  that some objects would trace the behaviour of 1 km bodies, some 10
  km, and so on.  However, in the absence of any collisional physics
  -- here, we use a simple hit-and-stick model with no fragmentation
  -- the small-radius planetesimals will not be replenished as they
  would be in a real system.  The absence of a collisional cloud may
  increase the eccentricity of the resulting planets as an object
  cannot be damped by its own debris, but this is probably negligible
  during the gas-rich phase when type I drag is strong and the embryos
  are on effectively circular orbits the majority of the time.

  Ultimately, it will be necessary to use a hybrid N-body/statistical
  scheme along the lines of \cite{brom06}, and we are currently
  exploring the incorporation of a variant of the semianalytic model
  of \cite{cham08} into the code.

  The choice of inner and outer edge for the initial solid material
  distribution in the N-body runs -- 1.0 AU and $\sim\!10$ AU -- may
  have changed the outcomes in two ways.  In the runs with low
  migration, some objects which should have been present in the
  interior will be missing.  However, this is seldom a significant
  amount of mass.  The only disc in which there was more than $4$\ME
  of material interior to 1 AU which was excluded was the $\fenh=10,
  \alpha=1.0$ case ($\sim\!8$\ME), which has very high migration;
  S22A, S36A, and S37A all fail to contain any objects inside of 1.5
  AU because they have all migrated away, and so very little material
  (if any) would have survived.  In the discs with the lowest
  migration rates, where $\fenh=3$, only $2.3$\ME \,of mass was missed
  when $\alpha=1.0$. (The flatter profiles have more of their mass at
  larger $r$ and so have even less initial mass inside of 1 AU.)
  Therefore the effects on the final results are limited:
  low-enhancement discs which have migration rates small enough that
  we should have included the innermost embryos also have too little
  mass in the region to be interesting, and high-enhancement discs
  which have enough mass to significantly affect the mass of a final
  planet have migration rates high enough that the missing material is
  long gone by the time the gas disc has dissipated.  The inner
  cut-off edge of 0.05 AU will also be responsible for removing some
  particles with small semimajor axis and high eccentricity, but there
  are very few of them.  On the outer edge, it is quite possible that
  a wider particle disc than 10 AU would have resulted in more planets
  in the high-enhancement runs and prevented the zero-planet outcomes
  (all of which were very high-migration configurations with
  $\fenh=10$ and $c_a$ = 1).  However, such high enhancements and
  migration rates are at the limits of plausibility to start with, and
  the initial conditions from the semianalytic model are questionable
  at high $\fenh$.  Most importantly, there is no reason to expect
  that the planets which would have been formed would be any larger
  than the ones already formed in the $\fenh = 10$ runs, as objects
  which migrate into the inner regions in such runs are often either
  well-separated or resonantly locked and therefore protected against
  most encounters.

  In our N-body model, the only effect that the gas disc has upon the
  particles is via prescription as a source of drag, whether
  aerodynamic or type I, and we did not include any disc physics
  involving gap opening.  This probably had little effect, as if we
  assume a (hydrodynamic-standard) constant $z_0/a$ ratio of $\simeq
  0.05$, and use the gravitothermal condition from \cite{crida06}, gap
  opening even in an inviscid disc requires $(3/4) (z_0/r_H) < 1$, or $r_H
  \simeq 3/4 H$, which gives a mass of $\simeq 53 \ME$ (with "gap
  opening" defined to be a 10\% perturbation in surface density).
  Even assuming a ratio of 1/2 sufficed to open a gap, we would need
  $\simeq17 \ME$.  Things become somewhat better in a flared disc
  (such as the one we actually used) with $z_0/a \propto a^{1/4}$,
  where if we assume no kinematic viscosity, an 8 \ME object could
  possibly start opening a gap at $\sim\!0.08$ AU, near the edge of
  our simulation region.  In practice, gap opening is unlikely to be a
  good explanation for the presence of the `lukewarm' end of the hot
  Neptune population unless it is substantially more efficient than
  currently thought.

  We also neglected the accretion of gas onto the embryo cores.  Given
  the low planet masses we obtained -- with median 1 \ME and maximum
  $\simeq8$ \ME -- the standard \cite{poll} model suggests that even
  our most massive planets would have accreted atmospheres of at most
  1-2 \ME.  This missing mass should not have affected the outcomes
  significantly, at least directly, as it would have changed the Hill
  radii by less than 10\%.  However, it is known \citep{inaba03} that
  growing embyros can have a significantly increased effective capture
  radius due to the energy loss suffered by planetesimals moving
  through the embryo atmosphere, which can help overcome the challenge
  of forming giant planet cores before the gas disc evaporates
  \citep{inaba03b}.  Would including this effect have made it possible
  to form substantially larger cores?

  This question should be considered in a larger context.  One feature
  common to many of the new mechanisms being discussed,
  atmosphere-enhanced capture radii included, is that they tend to
  increase the accretion rate at early times.  Partly this is due to a
  target bias on the part of scientists studying the middle and late
  stages of planet formation, as forming the cores of giant planets on
  appropriate time-scales is an important open problem even in the
  migration-free case, and so efforts have been concentrated in the
  direction of making large objects easier to form.  In the absence of
  migration, this is a net positive.  The situation is more
  complicated when migration is involved.  Increasing the accretion
  rate sounds appealing, but since the type I migration rate scales
  linearly with the mass, it will simultaneously decrease the
  migration time-scales, and do so at early times when the gas density
  is at its highest.  If anything, this will make the
  formation-to-migration time-scale problem worse, at least in cases
  where migration is significant.  \cite{komII} noted that for a fixed
  gas-to-dust ratio and gas dissipation time-scale, increases in
  initial surface density can result in decreases in final surface
  density.  \cite{cham08} finds that for a disc with alpha viscosity
  of 0.001, the maximum embryo mass increases as the disc mass
  increases to a high of $\simeq3 \ME$ at $0.05 \MSun$, and then falls
  off, so that disc masses of $0.03 \MSun$ and $0.07 \MSun$ both have
  maximum planet mass $0.5-0.6 \ME$.  Including fragmentation does
  raise the maximum mass reached over a wide range of disc mass
  ($0.03-0.10 \MSun$), but only to $\simeq4 \ME$.  The case with alpha
  viscosity of 0.01 resembles the case without migration below disc
  masses of $0.05 \ME$.  Only at very large disc masses, greater than
  0.10 \MSun, are Neptune-like masses recovered.
 
  Modifying the accretion rate seems unlikely to encourage
  neighbouring oligarchs to merge at late times, given their large
  separations.  These spacings are consequences of the high surface
  density of the discs as well as type I migration.  Enhanced capture
  radius mechanisms should have little direct effect on the giant
  impact phase, as a body needs to encounter roughly its own mass of
  gaseous material in the extended atmosphere to significantly affect
  its orbit.  For example, Earth-mass objects experiencing a
  near-collision will barely notice any atmosphere, however extended.
  In general, changes to the accretion rate in these oligarchic
  migration scenarios should move the window of source material which
  grows and migrates to the innermost regions, but will not increase
  the total mass by the factors of several needed to recover a
  planetary system resembling HD69830 (except possibly for massive
  discs).  That is, it should change where the planets come from, not
  how large they are.

  \subsection{Comparisons with previous work}
  \label{comparison}

  Our results differ significantly from previous results in the
  literature, chiefly in that we fail where they succeed.

  \cite{alib} succeeded in forming a system which resembles the
  HD69830 system very closely, by reducing the strength of the
  migration by a factor of 10 and considering only three cores which
  migrate and accrete (c.f.~\citealt*{growthofmig}).  As \cite{cham08}
  explains, the presence of interior embryos with a much shorter
  dynamical time means that inward-migrating cores do not encounter a
  pristine feeding zone, but one which is already substantially
  processed.  None of our migrating cores experienced growth
  resembling that in the \cite{alib} model.  \cite{payne2009} further
  find that aerodynamic drag can significantly decrease the
  planetesimal accretion rate in such scenarios, keeping the migrating
  core acting as a shepherd and not a predator.

  An alternative way of ensuring that a population of hot super-Earths
  or hot Neptunes can form and survive is to invoke an inner disc
  cavity generated by interaction between the disc and the stellar
  magnetosphere.  Such a model has been used to explain the existence
  of hot Jupiters \citep{linbod}, and was incorporated into the
  formation models of \cite{brun} and \cite{terq} (who also included
  tidal interaction with the central star).  \cite{brun} and
  \cite{terq} both succeeded in forming large short-period
  Neptune-mass (10--30 \ME) cores.  \cite{terq} form four objects of
  mass $\geq 8 \ME$, the largest being $12 \ME$, and \cite{brun} form
  five objects of mass $\geq 10 \ME$, with the largest being
  $\sim\!\!24 \ME$.  There are three main differences between their
  models and ours which likely account for the discrepancy.  First,
  both are relatively low-resolution: \cite{terq} used only 10
  Earth-mass bodies in all but one of their runs, and \cite{brun} used
  100 embryos of 0.5 \ME and 200 planetesimals of 0.1 \ME.  Second,
  and more importantly, it is unlikely that their initial masses and
  spacings can be recovered from a self-consistent oligarchic
  migration model: there is no plausible prior configuration which
  would evolve into (for example) 12 Earth-mass bodies spaced between
  0.1 and 1 AU.  Finally, and most importantly, in accordance with the
  cavity hypothesis both groups used a disc with an inner edge at
  either 0.05 AU or 0.10 AU, and we did not.  Some small toy
  simulations we performed (unpresented here) confirm that we can also
  build bodies of $\simeq15 \ME$ by doing so, but only relatively
  close to the boundary.  However, from general considerations the
  magnetospheric cavity in this model is expected to extend out to a
  distance of $\simeq 0.08$ AU for an assumed T Tauri star rotation
  period of 8 days \citep{linbod}, causing migration to halt at a
  distance of $\simeq 0.05$ AU from the star.  We simply note that
  there are a number of super-Earths and Neptune-mass extrasolar
  planets orbiting with significantly larger semimajor axes
  (e.g.~HD69830 d, or OGLE-05-169L b, at 2.8 AU) whose inward
  migration was probably not halted by the presence of an inner disc
  cavity. It is more likely that they, or their progenitor objects,
  were stranded at or near their current location when the gas disc
  dispersed.

  \cite{thommes07} and \cite{cham08} use semianalytic models, both
  incorporating atmosphere-enhanced capture radii and the latter
  fragmentation, and succeed in forming planets of $\ge 10 \ME$,
  although \cite{thommes07} produce far more than \cite{cham08}.  They
  both use an approach in which embryos are treated as discrete but
  non-interacting objects instead of as a continuous distribution
  (replacing the more Eulerian treatment of \citealt*{thommes03} and
  \citealt*{cham06} with a Lagrangian one).  In order to maintain the
  standard oligarchic separation of $\sim\!10$ Hill radii, they merge
  two embryos whenever their semimajor axes differ by less than 7 Hill
  radii.  As \cite{cham08} notes, this will result in missing some
  interesting inter-embryo dynamics such as migration convoys (groups
  of embryos migrating in mean-motion resonance, as in
  \citealt*{mcn05}; see also \citealt*{thommes05, papszu}), but it
  will also miss some larger and more fundamental dynamics such as
  deviations of the embryo behaviour from the semianalytic
  predictions, and introduce an unphysical orderly merger wave.

  \cite{cham08} performed a comparison between the results of his
  semianalytic model and one of the terrestrial-planet N-body
  integrations of \cite{mcn05}, and found that the N-body results had
  both larger median masses ($0.25 \ME$ vs. $0.16 \ME$) and larger
  median spacings ($\sim\!20$ Hill radii vs. 11).  Even in the
  original oligarchic model of \cite{ko98}, there is a weak dependence
  of $b$ on $M$ and $\Sigma$.  If in these scenarios the interembryo
  spacing grows to values much larger than $10$ while the objects stay
  on near-circular orbits, oligarchic growth can be quenched
  (e.g.~\citealt*{stab2, zhou07}).  As \cite{cham06} (\S8) notes, in
  order to maintain a constant spacing in $b$ within a fixed width,
  embryos must accrete 1/3 of their mass via embryo-embryo mergers.
  This accretion mode should vanish if the embryos are too
  well-separated (and too cold, but the models assume the embryos are
  always on circular orbits.)  Too large a spacing towards the end of
  the gas phase can also suppress the giant-impact phase almost
  entirely, requiring external sources of stirring to form large cores
  on reasonable time-scales.  Accordingly, if the interembryo spacing
  $b$ for large-mass bodies in these simulations is closer to $15-20$
  than $10$, then calibrating the (admittedly merely statistical)
  prescription for the embryo mergers to the no-migration,
  low-embryo-mass separation value of $10$ may not be appropriate.

  Despite the differences, we find some qualitative agreements with
  \cite{cham08}.  In his models with migration but without
  fragmentation (the closest match to our runs), he finds that the
  maximum mass of an object is maximized at $\simeq 3 \ME$ for a disc
  mass of $0.05 \MSun$, and that both higher and lower disc masses
  decrease this number (as we find, and \citealt*{komII} predicted).
  For his simulations which include both fragmentation and migration,
  as mentioned above, there is a wide range of disc masses ($0.03-0.10
  \MSun$) which produce roughly comparable maxima ($\simeq4 \ME$).
  The only simulations which succeed in getting cores larger than
  $10\ME$ either fail to include migration or include both
  fragmentation and migration with a disc mass of $\gtrsim 0.1 \MSun$,
  and it appears that the only Neptune-like object had a semimajor
  axis outside 2 AU.  This is in mild disagreement with
  \cite{thommes07}, who found 10 \ME objects were common, which
  \cite{cham08} argues is the result of their choices of a large alpha
  viscosity, 0.01, and a small planetesimal radius, 1 km, while
  neglecting fragmentation.

  We suspect that the solution to the short-period Neptune formation
  problem lies not in tweaking the accretion rate but in modifying the
  prescription for the interaction with the gas disc.  Obviously we
  need some kind of migration to build a short-period population, but
  what properties should it have?  Ideally, we would prefer a
  migration which: (1) has little to no effect at early times when the
  gas density is high, so that enhancing the surface density actually
  results in larger embryos and not more embryos lost, (2) when
  active, has a reduced rate from the nominal \cite{ttw} behaviour, as
  our $c_a=0.3$ runs performed noticeably better than our full
  migration runs, and (3) provides some mechanism to encourage large
  migrating embryos to merge instead of locking in convoys or
  stranding themselves at large separations from their neighbours,
  such as a variable migration direction.

  The most promising model of disc-induced migration which displays
  these characteristics is that presented by \cite{paard06, paard08},
  in which the migration of planets in radiatively inefficient discs
  was considered.  This model has the highly desirable property that
  migration is stopped, or even reversed, during early times when the
  disc is optically thick, but inward type I migration is recovered
  when the disc density decreases and the gas becomes optically
  thin. An alternative model for the modification of type I migration
  is stochastic migration induced by turbulent density fluctuations in
  the disc \citep{nelson04, nelson05}.  Although this does not appear
  to have the same nicely-tailored characteristics for solving the
  problem of short-period Neptune formation as the radiatively
  inefficient migration model, its random walk nature has the
  potential to deliver significant planetary material into the
  interior regions at late times.  
 
  \section{Conclusions}
  \label{section:conc}

  We performed 48 simulations of various oligarchic migration
  scenarios to determine whether the simplest standard approach can
  succeed in forming a population of short-period Neptune systems
  under common assumptions for the protoplanetary disc parameters.
  Multiple numerical techniques were applied: semianalytic techniques
  for the first 0.4 Myr, our new parallel multizone N-body code for
  the accretion phase while gas is present up to 6 Myr, and the more
  traditional SyMBA approach for the late stage to 100 Myr.

  We find that over a wide range of disc conditions, it is difficult
  to form planets of mass greater than $3-4 \ME$.  Our most successful
  runs involved $\sim$ 5 times the mass of the MMSN, surface density
  varying as $r^{-1/2}$, a disc decay time-scale of 1 Myr, and a
  migration efficiency of 0.3.  Our most common planet outcomes are of
  Earth-mass objects, with the terrestrial planets having ice
  fractions from 0.0 to 0.75 (the maximum possible in our
  simulations).  The larger objects have higher ice fractions, with
  the median being 0.60 for objects above 1 \ME, and 0.36 below.  In
  none of the cases did we succeed in forming an object of greater
  than 7.5 \ME inside 2 AU, much less inside 0.5 AU, and the total
  embryo mass remaining inside 2 AU was always less than 17 \ME.  The
  existence of an upper limit and the weak dependence on most
  parameters is in accordance with the predictions of \cite{komII},
  and in rough agreement with the predictions of \cite{cham08} except
  at large disc masses.  Nevertheless, we should be wary of making
  predictions based on these results regarding extrasolar planetary
  systems, as they entirely fail to reproduce the short-period Neptune
  planetary population that we know exists.

  Our failure can be compared to several previous successes in the
  literature, which either (1) adopt initial conditions which are not
  easily reconciled with an oligarchic growth picture, (2) use an
  inner edge to the migration (which is defensible but will have
  difficulty explaining more distant Neptunes), or (3) neglect
  inter-embryo dynamics and use an embryo merger condition which is
  calibrated to an effective inter-embryo separation (10 Hill radii)
  which is considerably smaller than we observe.

  Varying parameters which we kept constant such as the gas-to-dust
  ratio, incorporating additional accretion physics such as
  fragmentation, moving to extremely large disc masses or extremely
  weak migration, and simply performing more runs (and hoping for
  fortuitous late mergers), could possibly succeed in improving the
  maximum mass reached by a factor of two and therefore into the
  Neptune-like region.  It seems quite unlikely that they will
  increase the median mass enough to comfortably produce a population
  of multiple-planet short-period Neptune systems.  We conclude that
  forming a system like HD69830 will probably require a significant
  revision to the simple models explored here.

  If the standard oligarchy-plus-type-I-migration picture fails to
  reproduce the observed distribution of short-period exoplanets even
  at more extreme parameter values, then we must consider non-standard
  models.  Oligarchy is relatively well understood both analytically
  and numerically; by comparison type I migration is sensitive to
  poorly understood properties of the gas disc such as disc turbulence
  and local thermodynamic time-scales.  In a follow-up paper we
  consider the implications for hot exoplanet formation via oligarchy
  of alternate migration models (such as \citealt*{paard06}) which
  show some promise.

  \section*{Acknowledgments}
  
  DSM is very grateful for the warm hospitality of Makiko Nagasawa at
  the Edge Institute at the Tokyo Institute of Technology, where an
  early version of this work was presented.  We appreciate the careful
  review of the anonymous (but recognizable) referee.  The authors gratefully
  acknowledge the support of SFTC grant PP/D002265/1.  The simulations
  presented in this paper were performed using the QMUL HPC facilities
  funded through the SRIF initiative, without which the research would
  not have been possible.

  \label{lastpage}
  
\end{document}